\documentclass[11pt]{article}
\pdfoutput=1
\usepackage[a4paper]{geometry}
\usepackage{color}
\usepackage{float}
\usepackage[utf8]{inputenc}
\usepackage{a4wide}
\usepackage{amsmath,amssymb,amstext,amsthm,amssymb}
\usepackage{amsfonts}
\usepackage{framed}
\usepackage{graphicx}
\usepackage{subcaption}
\usepackage{bbold}
\usepackage{bbm}
\usepackage{slashed} 
\usepackage{tensor} 
\usepackage{braket} 
\usepackage{amstext}
\usepackage{array}
\usepackage{graphicx}
\usepackage{setspace}
\usepackage{hyperref}
\usepackage{overpic}
\usepackage{bbm}
\usepackage{cite}
\usepackage{lipsum}
\usepackage{tikz}

\newcommand{\be}{\begin{equation}}
\newcommand{\ee}{\end{equation}}
\newcommand{\bea}{\begin{eqnarray}}
\newcommand{\eea}{\end{eqnarray}}

\begin{document}

\sloppy
\pagenumbering{arabic}
\thispagestyle{empty}
\begin{center}
{\huge \bf Gauge and gravitational instantons:}\\[.5cm] 
{\huge \bf From 3-forms and fermions to}\\[.3cm] 
{\huge \bf Weak Gravity and flat axion potentials}\\
\vspace{1.5cm}
{\large Arthur Hebecker and Philipp Henkenjohann}\\
\vspace{0.5cm}
\textit{
\vspace{0.3cm}
Institut f\"ur Theoretische Physik, Universit\"at Heidelberg, \\
Philosophenweg 19, 69120 Heidelberg, Germany}\\
\vspace{1cm}
June 18, 2019

\vspace{1cm}
\textbf{Abstract}\\
\end{center}
We investigate the role of gauge and gravitational instantons in the context of the Swampland program. Our focus is on the global symmetry breaking they induce, especially in the presence of fermions. We first recall and make more precise the description of the dilute instanton gas through a 3-form gauge theory. In this language, the familiar suppression of instanton effects by light fermions can be understood as the  decoupling of the 3-form. Even if all fermions remain massive, such decoupling may occur on the basis of an explicitly unbroken but anomalous global symmetry in the fermionic sector. This should be forbidden by quantum gravity, which leads us to conjecture a related, cutoff-dependent lower bound on the induced axion potential. Finally, we note that the gravitational counterpart of the above are K3 instantons. These are small fluctuations of Euclidean spacetime with K3 topology, which induce fermionic operators analogous to the 't Hooft vertex in gauge theories. Although Planck-suppressed, they may be  phenomenologically relevant if accompanied by other higher-dimension fermion operators or if the K3 carries appropriate gauge fluxes.

\vspace{0.5cm}
\newpage

\section{Introduction and overview}
\subsection{Fermions in the 3-form description}

In this paper, we investigate the effective description of gauge or gravitational instantons by 3-form gauge theories, focusing in particular on the effect of fermions on this effective theory. Furthermore, we discuss how the interplay of fermionic operators and instantons affects axionic shift symmetries. If exact, such global symmetries should be in the swampland and we propose a lower bound on axion masses to quantify the minimal strength of symmetry breaking. Such a bound can then be used to derive constraints on the fermionic operators mentioned above. 

Gauge theories of 3-form potentials have been discussed since a long time \cite{Aurilia:1978dw}. It is also well-known that the Chern-Simons 3-form of a non-Abelian gauge theory transforms under the non-Abelian gauge transformation exactly like a fundamental 3-form gauge potential \cite{Luscher:1978rn}. A similar argument can be made for gravity. Taking this seriously, one can use the 3-form gauge theory as an effective description of Yang-Mills (YM) theory at low energies. This has been discussed in \cite{DiVecchia:1980yfw} in the context of chiral perturbation theory of QCD and more generally in \cite{Dvali:2013cpa,Dvali:2016uhn,Dvali:2016eay,Dvali:2017mpy}.

Our first, simple, technical point in this paper is to apply this description to the case of a Higgsed YM theory at energies below the symmetry breaking scale. In this regime, the instanton gas is dilute and a quantitatively controlled analysis is possible. As a result, the effective 3-form gauge theory description can be rigorously established as long as the source term $\theta$ (used for probing the theory through the coupling $\theta\,$tr$(F\tilde{F})$) is small. We will rely on this controlled model of a dilute instanton gas and its 3-form description in what follows.

We are particularly interested in the 3-form description of YM theories with instantons and massless fermions~\cite{Dvali:2013cpa,Dvali:2016uhn, Dvali:2016eay, Dvali:2017mpy}. Especially the idea that this may imply fermion condensates independently of confinement and that small fermion masses may be generated through gravitational instantons are intriguing.

A crucial assumption for this line of reasoning is that the effective 3-form description of YM theory with massless fermions includes a massless pseudoscalar. Indeed, in QCD this is the familiar $\eta'$ meson. This scalar may be dualized into a 2-form which then gauges the effective 3-form, making it massive \cite{Dvali:2005an}. As a result, one obtains a very reasonable effective description for how fermions remove instanton effects. However, as we point out, a different option arises in our case of a Higgsed YM system: Light fermions may suppress the gauge coupling of the effective 3-form, completely decoupling it in the massless limit. This interpretation is supported, in our calculable setting, by the fact that no evidence of the massless scalar can be found. Thus, the effective description of how fermions remove instanton effects may change depending on the diluteness of the gas.

\subsection{Fermions and axion potentials}

We go on to study Higgsed YM theories which are coupled to an axion. Gauge instantons then generically induce an axion potential $\propto \text{e}^{-S}$ where $S$ denotes the instanton action. According to the Weak Gravity Conjecture (WGC) for axions, $S$ is bounded from above as $S\lesssim M_\text{P}/f$, with $f$ the axion decay constant \cite{ArkaniHamed:2006dz}.\footnote{
To 
be precise, most naively the WGC constrains the mass (action) of a charged object to be small at weak coupling. For axions, this is the large-$f$ regime, $M_\text{P}/f\ll 1$. We assume here that the bound $S\lesssim M_\text{P}/f$ is (or is also) valid at small $f$, which is our main case of interest in this paper.
}
However, a priori the WGC does not make a claim about the overall size of the axion potential. Thus, the potential could be small because $\text{e}^{-S}$ comes with a small prefactor \cite{delaFuente:2014aca,Hebecker:2018ofv,Shiu:2018wzf,Shiu:2018unx}. In the following we want to take the bound on $S$ seriously and focus on the prefactor.

The smallness of the instanton prefactor can, in particular, be due to the presence of light fermions with mass $m$. Indeed, the prefactor scales as $(m/v)^{N_\text{f}}$, with $v$ the Higgs scale and $N_\text{f}$ the number of flavors. Obviously, for $m\rightarrow0$ the potential vanishes identically and a global symmetry involving a shift in the axion and anomalous U(1) rotations of fermions emerges. Unless that symmetry is broken, for example by additional fermion interactions, this is inconsistent with quantum gravity expectations \cite{Abbott:1989jw,Coleman:1989zu,Kallosh:1995hi,Banks:2010zn}. 

Intriguingly, a similar phenomenon can be observed in a model in which all fermions remain heavy. As we will explain, this is achieved by shift-symmetry-preserving Yukawa interactions, which provide effective mass operators in addition to the hard mass terms. In fact, such a structure arises in the Standard Model if an axion coupling to tr$_{SU(2)}F\tilde{F}$ is introduced~\cite{Nomura:2000yk}. Again, once the hard masses are taken to zero, the axion potential vanishes due to the emergence of a global symmetry. This shows that the problematic feature of the theory is not massless fermions but rather the presence of a global symmetry. Note also that in this class of models one can apparently take the massless-axion limit without tempering with any other IR degrees of freedom.

One could restore consistency with quantum gravity by simply claiming that flat axion potentials are in the swampland, thereby excluding such models. However, there are many counterexample in string theory: Calabi-Yau compactifications of type II strings lead to 4d $\mathcal{N}=2$ supergravity models with perfectly flat moduli spaces, including axionic directions. Of course, in this case we expect that the global axionic symmetries are broken by higher-dimension operators involving fermions. The latter stay massless due to supersymmetry and a low-energy theory with a global shift symmetry is not realized even at arbitrarily low energy.

We note that much more work related to the WGC, especially the WGC for axions, has recently been done (see e.g.~\cite{Cheung:2014vva,delaFuente:2014aca,Rudelius:2015xta,Montero:2015ofa,Brown:2015iha,Bachlechner:2015qja,Hebecker:2015rya,Brown:2015lia,Junghans:2015hba,Heidenreich:2015wga,Palti:2015xra,Heidenreich:2015nta,Kooner:2015rza,Kaloper:2015jcz,Kappl:2015esy,Choi:2015fiu,Kaplan:2015fuy, Ibanez:2015fcv,Heidenreich:2016aqi,Klaewer:2016kiy,Hebecker:2017wsu,Lee:2018spm,Hebecker:2018yxs,Lee:2019tst,Marchesano:2019ifh,Grimm:2019wtx})\footnote{See \cite{Palti:2019pca} for a recent review on the WGC and, more generally, on the swampland paradigm.}. A lot of this is in the context of trying to avoid or defend the bound on $S$. An imortant ingredient is trying to break the microscopic connection between $f$ and $S$~\cite{Kim:2004rp}, with recent concrete realizations including in particular~\cite{Hebecker:2015rya, Choi:2015fiu, Kaplan:2015fuy, Hebecker:2018yxs}. Moreover, it has been argued that  wormholes generate a potential for axions \cite{Rey:1989mg} (see also \cite{Montero:2015ofa,Alonso:2017avz,Hebecker:2018ofv}) which, however, we will not discuss in the following. Phenomenological consequences of axion potentials for small $f$ are discussed in \cite{Alonso:2017avz} and \cite{Hebecker:2018ofv}. In spirit close to our discussion is a proposal in \cite{Moritz:2018sui} which constrains the relative size of any corrections (not just instantons) to axion potentials. The strong form of this constraint requires that there always exists a dominant contribution to the axion potential that has a sub-Planckian periodicity (corresponding to $f<M_\text{P}$). Still, this constraint does not make a claim about the absolute size of axion potentials.

\subsection{Flat axionic potentials and the Swampland}

Ultimately we are interested in finding a general lower bound on axion potentials, thereby quantifying to which extent an approximate global axionic shift symmetry is compatible with quantum gravity. To do so, we try to invoke the WGC for the effective 3-form theory encoding the crucial instanton effects. The magnetic version of the WGC for 3-forms bounds the cutoff $\mu$ of the 3-form theory according to $\mu\lesssim(\Lambda^2M_\text{P})^{1/3}$, where $\Lambda^2$ denotes the 3-form gauge coupling. Unfortunately, the restriction of our effective 3-form theory to small $\theta$ translates to a restriction to energy scales below $\Lambda$. Thus, we never enter the interesting regime above $(\Lambda^2M_\text{P})^{1/3}$ where WGC constraints would apply.

To improve on this, we argue for the existence of an extended, highly non-linear 3-form description which is valid beyond $\theta\ll1$. This enhanced range of validity allows for a cutoff which can be larger than $\Lambda$ and is given by the mass scale of the lightest massive degrees of freedom in the UV, for example the mass of light fermions or the Higgs scale. In contrast to the original 3-form description, this new 3-form theory is severely constrained by the WGC. For example, it would parametrically constrain the Higgs scale $v$ of Higgsed YM models with gauge coupling $g$ according to $v\lesssim M_\text{P}\exp(-1/g^2)$. This would put weakly coupled YM theories which are Higgsed at a high scale into the swampland. We feel that this is too strong a statement to be taken seriously and conclude that the WGC for 3-forms is probably only valid for the canonical quadratic 3-form theory.

Thus, while the 3-form-WGC-approach fails, we still expect some lower 
bound on generic axion potentials \mbox{$V(\phi)=-V_0\text{e}^{-S}\cos\phi$} to exist. We are inspired by the axionic WGC, but we also know from the $\mathcal{N}=2$ SUGRA example that the bound should disappear as the cutoff is taken to zero. Guided by this as well as by simplicity and consistency with known examples we hence propose that 
\begin{equation}
 V_0\text{e}^{-S}\gtrsim\mu^\alpha M_\text{P}^{4-\alpha}\exp(-cM_\text{P}/f)\,,
\end{equation}
where $\alpha$ and $c$ parametrize our ignorance of the exact bound, $f$ denotes the axion decay constant and $\mu$ is the cutoff of the low energy theory that exclusively describes the axion. From this one can easily derive a bound on the axion mass $m_\phi$:
\begin{equation}
 m_\phi\gtrsim\frac{M_\text{P}^2}{f}\left(\frac{\mu}{M_\text{P}}\right)^{\alpha/2}\exp\left(-c\frac{M_\text{P}}{2f}\right)\,.
\end{equation}

We have already mentioned that we expect global axionic shift symmetries to be broken by appropriate fermion operators. If the corresponding fermions are not massless, they can be integrated out, thereby inducing an axion potential. Our proposed bound on such potentials then translates to a lower bound on fermion operators. We apply this approach to fermion masses in Higgsed YM theory with and without additional mass contributions from Yukawa couplings. In both cases we find that fermion masses $m$ are parametrically bounded by 
\begin{equation}
 m\gtrsim\exp\left(-c\frac{M_\text{P}}{f}\right)v
\end{equation}
for $\alpha=4$ with the caveat that in the case without Yukawa couplings this holds only for more than four fermion flavors.

\subsection{Gravitational instantons}

Finally, we argue for the possibility of gravity-induced fermion interactions via gravitational instantons. To appreciate our argument, recall that the axial U(1) symmetry of gauged massless fermions is anomalously broken according to $\partial_\mu J^\mu_\text{A}\propto\text{tr}(F\tilde{F})$, where $J^\mu_\text{A}$ is the axial U(1) current. In the presence of gauge instantons, the spacetime integral of the topological density $\text{tr}(F\tilde{F})$ is non-trivial. This implies the existence of an effective fermion interaction that explicitly breaks the axial U(1). This instanton-induced fermion interaction, also called 't~Hooft interaction, can be determined explicitly in the dilute gas approximation \cite{tHooft:1976snw,tHooft:1986ooh}.

The same logic can be applied to pure gravity. There we have $\partial_\mu J^\mu_\text{A}\propto\text{tr}(R\tilde{R})$ with the right hand side being a gravitational topological term. To the best of our knowledge, K3 is the only compact manifold non-trivially contributing to this term \cite{Hawking:1979zw,Hawking:1979zs}. We describe how a K3 manifold can be glued into flat $\mathbb{R}^4$, such that it can be considered a local fluctuation of it. This is analogous to the localized field strength of a gauge instanton and also implies the gravitational analogues of 't Hooft fermion interactions. Using the dilute gas approximation we naively estimate the strength of these interactions and find that the associated energy scale is, in the case of the SM, of the order of $10^{17}$ GeV. We therefore conclude that they are phenomenologically irrelevant. However, we point out that this may be circumvented by combining the K3-instanton-induced interactions with other higher-dimension operators.

The rest of the paper is organized as follows: In Sect.~\ref{sec::3forms}, we recall and develop some basic ideas about 3-form gauge theories. We then use these theories as effective descriptions of gauge instantons (with and without fermions) in Sect.~\ref{sec::3eft}. This setting and other approaches are employed in Sect~\ref{sec::swamp} to discuss possible Swampland constraints on the achievable flatness of axion potentials. Section~\ref{sec::grav} analyzes analogous effects of gravitational instantons and some Conclusions are drawn in Sect.~\ref{sec::conc}. 

\section{The physics of massless and massive 3-forms}

\label{sec::3forms}

In this section we collect some results about 3-form gauge theories \cite{Aurilia:1978dw, Luscher:1978rn, DiVecchia:1980yfw, Dvali:2013cpa} (see Appendix A for details and derivations). In particular, we argue for the equivalence with the dual $(-1)$-form description. We also show how the force between domain walls can be used to follow the transition between Coulomb and Higgs phase. This quantifies features discussed in~\cite{Dvali:2005an}. 

The free theory is defined by the Euclidean action
\begin{equation}
 S_\text{E}[A_3,\theta]=\int_{M_4}\left(\frac{1}{2\Lambda^4}F_4\wedge*F_4-\text{i}\theta F_4\right)\,, \label{actionpure}
\end{equation}
where $F_4=\text{d}A_3$ is the field strength, $\Lambda^2$ is the gauge coupling, and $M_4$ is the 4d Riemannian manifold on which the theory lives. If $M_4$ is compact, the $F_4$-flux on it is quantized and one can dualize the partition function based on (\ref{actionpure}) in terms of a sum over discrete values of $F_0$ (see Appendix \ref{app::pure}). Explicitly, 
\begin{equation}
 Z[\theta]= \int {\cal D} A_3 \exp(-S_E[A_3,\theta])=C \sum_n\exp\left(-\frac{\Lambda^4}{2}\int_{M_4}(\theta+2\pi n)^2*1\right)\,,
\end{equation}
where $C$ is a normalization constant. This partition function is invariant under the shift $\theta\rightarrow\theta+2\pi$  and we can hence view $\theta$ as a periodic variable which takes values in the range $[-\pi,\pi)$. Using this we conclude that for constant $\theta$ the term with $n=0$ corresponds to the lowest energy state. In the limit $M_4\to \mathbb{R}^4$ this term dominates:
\begin{equation}
Z[\theta]\propto\exp\left(-\frac{\Lambda^4}{2}\int \theta^2*1\right)\,. \label{partfuncpure}
\end{equation}
We also note that, for compact space and non-compact (Euclidean) time, i.e. for  $M_4=\mathbb{R}\times M_3$, the theory clearly represents a non-trivial quantum mechanical system: The fundamental degree of freedom can be characterized by $\int_{M_3}A_3$. This system corresponds to a quantum particle on a circle. From now on we will, however, take $M_4=\mathbb{R}^4$ for simplicity. 

Let us now introduce Cartesian coordinates $(x,y,z,t)$ and choose the source $\theta$ such that it represents two parallel domain walls localized at $x=a$ and $x=b>a$: 
\begin{equation}
 \theta(x)=
 \begin{cases}
  \theta_1 & \text{for } x\le a \\
  \theta_2 & \text{for } a < x < b \\
  \theta_1 & \text{for } b\le x
 \end{cases}\,.
\end{equation}
They are subject to a force per unit area,
\begin{equation}
 f^{(a)}=-f^{(b)}=\frac{\Lambda^4}{2}(\theta_2^2-\theta_1^2)\,, \label{forcepure}
\end{equation}
which is independent of the distance $b-a$. This is expected since one has no propagating degrees of freedom. Instead, the force is due to a constant background field strength $F_4$ which is different between the walls and outside. 

Next we consider the coupling of a dynamical scalar $\phi$ with mass $m$ to the 3-form:
\begin{equation}
 S_\text{E}[A_3,\phi,\theta]=\int_{M_4}\left(\frac{1}{2\Lambda^4}F_4\wedge*F_4-\text{i}(\theta+\phi) F_4+\frac{f^2}{2}\text{d}\phi\wedge*\text{d}\phi+\frac{1}{2}m^2f^2\phi^2*1\right)\,.
\end{equation}
Here $f$ determines the normalization of $\phi$. If $m=0$, one may take the scalar to be periodic such that $f$ becomes its axion decay constant. The action above is then dual to the situation where a 2-form is gauged by a 3-form as discussed in \cite{Dvali:2005an}.\footnote{
This 
is the gauge-field-theoretic description~\cite{Kaloper:2008fb, Kaloper:2011jz} of axion monodromy inflation~\cite{Silverstein:2008sg, McAllister:2008hb}, recently revived in the context of $F$-term axion monodromy~\cite{Marchesano:2014mla, Blumenhagen:2014gta, Hebecker:2014eua}. 
} 
For us, $m$ is a convenient parameter to switch this gauging on and off. Indeed, if $m\ne0$ the field $\phi$ disappears in the IR and $A_3$ is not Higgsed.\footnote{
Note 
that a non-zero $m$ may even be made consistent with a fundamentally axionic nature of $\phi$: All one needs is to interpret the effect of the gauging of $\phi$ by a further 3-form (which has been integrated out) as an effective monodromy or mass parameter.
}

Integrating out $A_3$ gives
\begin{equation}
Z[\theta]\propto\int\mathcal{D}\phi\exp\left[-\int_{M_4}\left(\frac{\Lambda^4}{2}(\phi+\theta)^2*1+\frac{f^2}{2}\text{d}\phi\wedge *\text{d}\phi+\frac{1}{2}m^2f^2\phi^2*1\right)\right] \label{partfuncscalar1}
\end{equation}
which, upon carrying out the $\phi$ integration, simplifies to
\begin{equation}
 Z[\theta]\propto\exp\left(-\frac{\Lambda^4}{2}\int_{M_4}\theta(x)\frac{\Box-m^2}{\Box-M^2}\theta(x)*1\right)
\end{equation}
with $M^2=m^2+\Lambda^4/f^2$. Also the force per area is altered:
\begin{equation}
  f^{(a)}=-f^{(b)}=\frac{\Lambda^4}{2}\left[\frac{m^2}{M^2}(\theta_2^2-\theta_1^2)+\left(1-\frac{m^2}{M^2}\right)(\theta_2-\theta_1)^2\text{e}^{-M(b-a)}\right]\,. \label{forcescalar}
\end{equation}
The additional term exponentially decays with the distance of the two domain walls and indicates the presence of a propagating degree of freedom with mass $M$. The effect of the constant background field strength is also still present, but it is now suppressed by $m^2/M^2$. At $m=0$, the 3-form theory is Higgsed and this long-distance effect disappears.

\section{3-form gauge theory as effective field theory of instantons}

\label{sec::3eft}

In the late 70s it has been noted that YM theory always contains a 3-form $C_3$ that inherits a corresponding gauge transformation, $C_3\rightarrow C_3+\text{d}\Omega_2$, from the original non-Abelian gauge symmetry. It therefore can be considered a proper gauge 3-form \cite{Luscher:1978rn}. More recently it was argued that this 3-form may provide an alternative description of the Peccei-Quinn solution of the strong CP problem in terms of a 3-form which gauges the 2-form dual to the axion \cite{Dvali:2005an}. This logic and its implications have been developed further in \cite{Dvali:2013cpa,Dvali:2016uhn}.

In this section we want to analyze the relation between YM theory and 3-form gauge theory more systematically. To do so we focus on the calculable case of a weakly coupled Higgsed YM theory such that we can employ the dilute instanton gas approximation in our computations. We find that the instanton induced correction to the vacuum energy can be effectively described by a pure 3-form gauge theory at small $\theta$-angle and below the Higgs scale. In the presence of light fermions this effective description remains a good approximation below the fermion mass.

\subsection{Pure Yang-Mills theory}

\label{subsec::pureym}

Let us first consider pure YM theory with Euclidean action 
\begin{equation}
 S_\text{E}[A_1,\theta]=\int\left(\frac{1}{2g^2}\text{tr}(F_2\wedge*F_2)-\frac{\text{i}\theta}{8\pi^2}\text{tr}(F_2\wedge F_2)\right)\,, \label{actioninst}
\end{equation}
where $A_1$ is the Lie-algebra-valued gauge potential and $F_2=\text{d}A_1$. As is well known, \mbox{$\text{tr}(F_2\wedge F_2)/(8\pi^2)$} is a total derivative, i.e.~it can be written as the exterior derivative of a 3-form $C_3$. This is the  proper 3-form gauge potential mentioned in the introduction to this section \cite{Luscher:1978rn}. $g$ denotes the gauge coupling and $\theta$ is again an arbitrary external source. For $\theta=\text{const.}$ we may identify it with the usual $\theta$-parameter of YM theory. In order to be able to deal with this theory computationally we assume the gauge symmetry to be broken spontaneously at a scale $v$ and take the running gauge coupling to be small at this scale: $g(v)\ll 1$. All instantons larger than $v^{-1}$ are then cut off and the dilute instanton gas approximation is valid. In this case one can integrate out the gauge field and obtains the partition function
\begin{equation}
Z[\theta]\propto\exp\left(2Kv^4\text{e}^{-S}\int d^4x\cos\theta\right)\,, \label{partfuncinstfull}
\end{equation}
where $S=8\pi^2/g(v)^2$ denotes the instanton action and $K\propto S^4$ (see Appendix \ref{app::b} for details). For small $\theta$ this reduces to
\begin{equation}
 Z[\theta]\propto\exp\left(-Kv^4\text{e}^{-S}\int d^4x\,\theta^2\right)\,, \label{partfuncinst}
\end{equation}
which is exactly the same as (\ref{partfuncpure}) for the pure 3-form gauge theory if we set
\begin{equation}
 \Lambda^4=2Kv^4\text{e}^{-S}\,. \label{3formcouppure}
\end{equation}By comparing the actions (\ref{actionpure}) and (\ref{actioninst}) we see that $\theta$ generates the same correlation functions for $F_4$ in the 3-form gauge theory as for $\text{tr}(F_2\wedge F_2)/(8\pi^2)$ in the gauge theory. Also the forces on domain walls will obviously be the same. Hence we have established the pure 3-form gauge theory as an EFT of Higgsed YM theory at energies below the symmetry breaking scale $v$.

\subsection{Yang-Mills theory with fermions}

\subsubsection{Comparison at the 1-instanton level}

Next we add fermions with mass $m$ to the Higgsed YM theory. For simplicity we add only one fermion field in the fundamental representation of the gauge group:\footnote{For the sake of simplicity we have not included the Higgs sector in the action which is, nevertheless, always implicitly assumed to be present.}
\begin{equation}
  S_\text{E}[A_1,\psi,\theta]=\int\left(\frac{1}{2g^2}\text{tr}(F_2\wedge*F_2)-\frac{\text{i}\theta}{8\pi^2}\text{tr}(F_2\wedge F_2)+\overline{\psi}(\hat{\gamma}_\mu\hat{D}_\mu+m)\psi*1\right)\,.
\end{equation}

If $m\gg v$, we can first integrate out the fermions finding an effective gauge theory action with additional terms suppressed by powers of $v/m$ \cite{Vainshtein:1981wh}. Ignoring these small corrections at and below the Higgsing scale $v$ we can continue to use the analysis of Subsection \ref{subsec::pureym}.

By contrast, light fermions (with mass $m\lesssim v$) have a significant effect. To see this, we integrate out the gauge field first and find the effective action for the fermions in a background of a dilute instanton gas. The corresponding calculation has been done by 't Hooft \cite{tHooft:1976snw,tHooft:1986ooh} and leads to the following partition function:
\begin{equation}
   Z[\theta]=\int\mathcal{D}\overline{\psi}\mathcal{D}\psi\exp\left[-\int d^4x\left(\overline{\psi}(\hat{\gamma}_\mu\partial_\mu+m)\psi+\kappa v\text{e}^{-S}(\overline{\psi}P_\text{L}\psi\text{e}^{\text{i}\theta}+\overline{\psi}P_\text{R}\psi\text{e}^{-\text{i}\theta})\right)\right]\,. \label{partfuncthooft}
\end{equation}
$\kappa$ is some constant and $P_\text{L/R}$ is the left- and right-handed projection operator, respectively. Note that the instanton induced 2-fermion interaction corresponds to the well-known 't Hooft determinant for one flavor and is suppressed by the instanton action via $\text{e}^{-S}$. In the process of integrating out the fermions these interactions will give rise to loop corrections to the effective action which are suppressed by powers of $\text{e}^{-S}$ and correspond to multi-instanton effects. Therefore it is possible to view the effective action as a power series in this suppression factor.

For now let us ignore these loop corrections and calculate the partition function to leading order in $\text{e}^{-S}$, i.e.~at the 1-instanton level, which can be done exactly \cite{tHooft:1976snw}. The result is
\begin{equation}
  Z[\theta]\propto\exp\left(2K'v^4\frac{m}{v}\text{e}^{-S}\int d^4x\cos\theta\right)
\end{equation}
and reduces to
\begin{equation}
  Z[\theta]\propto\exp\left(-K'v^4\frac{m}{v}\text{e}^{-S}\int d^4x\,\theta^2\right) \label{partfuncinstfermi1}
\end{equation}
for small $\theta$. Fortunately, this exactly coincides with (\ref{partfuncinst}) up to a suppression factor $m/v$ and therefore we can once again apply the logic of Subsection \ref{subsec::pureym} to conclude that, at leading order in $\text{e}^{-S}$, Higgsed YM theory with a light fermion is, at energies below the fermion mass $m$, effectively described by a 3-form gauge theory (\ref{actionpure}) with
\begin{equation}
\Lambda^4=2K'v^4\frac{m}{v}\text{e}^{-S}\,. \label{lambdainstfermi1}
\end{equation}
For the sake of completeness let us let us also give the corresponding result for $N_\text{f}$ fermion flavors with mass $m$ (cf. Appendix \ref{app::b}):
\begin{equation}
\Lambda^4=2K'v^4\left(\frac{m}{v}\right)^{N_\text{f}}\text{e}^{-S}\,. \label{lambdainstfermiN}
\end{equation}
This shows that in the limit of massless fermions, $m\rightarrow 0$, the 3-form gauge coupling $\Lambda^2$ vanishes or, in other words, the 3-form becomes non-dynamical. At the same time, the cutoff of the effective 3-form theory goes to zero of course. This consistently reproduces the fact that the $\theta$-parameter of YM theory becomes unphysical and instantons are suppressed in the presence of massless fermions.

Let us give an intermediate summary and make an observation which we find interesting: We are considering a YM theory that is Higgsed at a scale $v$ and contains light fermions of mass $m$ below $v$. We may assume $m\ll v$, such that an EFT at scale $\mu$ with $m\ll\mu\ll v$ can be defined. In this EFT, the massive gauge bosons have been integrated out such that we are dealing with a purely fermionic theory. In addition to the kinetic and mass term, these fermions are subject to the famous, instanton-induced 't Hooft interaction. Next, we may also integrate out the fermions (at the 1-instanton level) to obtain the EFT relevant at scales below $m$. We argued that this is a massless 3-form gauge theory. The only assumptions were small $\theta$ and that higher-order corrections in $\text{e}^{-S}$ do not modify \eqref{partfuncinstfermi1} significantly. The interesting implication of this is that the low-energy limit of a fermion theory with 't Hooft interactions is provided by a 3-form theory. Note that this 3-form has a priori nothing to do with the 3-form $C_3\propto\text{tr}(\text{d}A\wedge A+A\wedge A\wedge A)$ present in the original YM theory.

\subsubsection{Multi-instanton effects}

\begin{figure}
  \begin{subfigure}[c]{0.49\textwidth}
  \center
    \begin{tikzpicture}
\draw (0,0) circle (1);
\draw (0,-1) ++(-45:0.2) -- +(135:0.4);
\draw (0,-1) ++(45:0.2) -- +(225:0.4);
\draw (0,1) -- +(135:0.15);
\draw (0,1) -- +(225:0.15);
\end{tikzpicture}
\caption{}
    \label{loopdiag1}
  \end{subfigure}
  \begin{subfigure}[c]{0.5\textwidth}
  \center
    \begin{tikzpicture}
\draw (0,0) circle (1);
\draw (0,-1) ++(-45:0.2) -- +(135:0.4);
\draw (0,-1) ++(45:0.2) -- +(225:0.4);
\draw (0,1) ++(-45:0.2) -- +(135:0.4);
\draw (0,1) ++(45:0.2) -- +(225:0.4);
\draw (1,0) -- +(45:0.15);
\draw (1,0) -- +(135:0.15);
\draw (-1,0) -- +(225:0.15);
\draw (-1,0) -- +(315:0.15);
\end{tikzpicture}
\caption{}
    \label{loopdiag2}
  \end{subfigure}
  \caption{The two diagrams are shown which contribute at leading (a) and sub-leading order (b) in the small parameter $\text{e}^{-S}$ to the vacuum energy. Crosses correspond to the 2-fermion interaction in \eqref{partfuncthooft}, which is proportional to $\text{e}^{-S}$.}
  \label{loopdiag}
\end{figure}
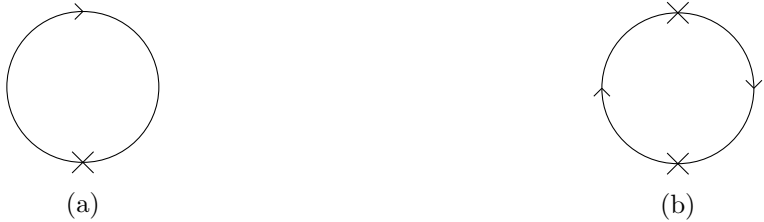

Let us now consider the next-to-leading order corrections due to the instanton induced fermion interaction. We want to clarify whether they can significantly affect our 3-form EFT in the relevant energy range $0\le\mu\le m$. To do so, the corrections to the force between two domain walls are calculated in Appendix \ref{app::b2}. The diagrams contributing to the energy are shown in Figure \ref{loopdiag}. Upon differentiation of this energy with respect to the distance between the domain walls, $b-a$, we find the force density
\begin{align}
 f^{(a)}=-f^{(b)}=\dots & +2\kappa^2v^4\left(\frac{m}{v}\text{e}^{-S}\right)^2\left(-\int\frac{d^3p}{(2\pi)^3}\frac{1}{\omega_p^3}(\theta_2^2-\theta_1^2)\right. \nonumber \\ 
 & \left.+\int\frac{d^3p}{(2\pi)^3}\frac{1}{\omega_pm^2}\text{e}^{-2\omega_p(b-a)}(\theta_2-\theta_1)^2\right)+\mathcal{O}\left(\frac{m}{v}\text{e}^{-S}\right)^3\,. \label{forcefermions}
 \end{align}
The dots $\dots$ denote the leading order contribution (\ref{forcepure}) with (\ref{lambdainstfermi1}) which is reproduced by the leading diagram Figure \ref{loopdiag1} as expected. The first term in the brackets corrects the leading order result and hence contributes to the gauge coupling $\Lambda^2$ of the effective 3-form theory. We expect that, at the intuitive level, this corresponds to a renormalization of the fermion mass $m$, possibly due to non-perturbative effects, such that \eqref{lambdainstfermi1} remains true when used with the appropriately renormalized mass. The second term, however, is finite and contains a non-trivial dependence on the distance between the domain walls.

We can rewrite this second term using the modified Bessel function $K_1$ of the second kind:
\begin{equation}
\int\frac{d^3p}{(2\pi)^3}\frac{1}{\omega_pm^2}\text{e}^{-2\omega_p(b-a)}=\frac{1}{2\pi^2}\frac{K_1(2m(b-a))}{2m(b-a)}\,. \label{forcefermionsbessel}
\end{equation}
For very small distances, $m(b-a)\ll1$, as well as very large distances, $m(b-a)\gg1$, this can be approximated as
\begin{equation}
  \frac{1}{2\pi^2}\frac{K_1(2m(b-a))}{2m(b-a)}=
 \begin{cases}
  \frac{1}{2\pi^2}\frac{1}{(2m(b-a))^2}+\mathcal{O}(1) & \text{for }m(b-a)\ll1 \\
  \left[\frac{1}{(4\pi m(b-a))^{3/2}}+\mathcal{O}\left(\frac{1}{(m(b-a))^2}\right)\right]\text{e}^{-2m(b-a)} & \text{for }m(b-a)\gg1
 \end{cases}
 \,. \label{forcecorrection}
\end{equation}
Note that in the limit $m\rightarrow0$ the force becomes exactly proportional to $1/(b-a)^2$.\footnote{In this limit one must not forget about the factor $m^2$ in front of the bracket in (\ref{forcefermions}).} Furthermore, our loop calculation indicates that it is due to the exchange of two fermions. Indeed, if the force were due to the exchange of one massless scalar, it should be independent of $(b-a)$, consistent with the Coulomb law in co-dimension one. By contrast, our observed $1/(b-a)^2$-behavior is consistent with the faster decay of a force originating from multi-particle exchange.

In \cite{Hsu:1992tg} it has been argued that the exchange of two massless fermions leads to a force that falls off as $1/r^6$ which, at first sight, seems to be in contradiction with the $1/r^2$ behavior found above. However, in our scenario the fermions interact with a classical source while in \cite{Hsu:1992tg} they mediate a force between two fermions via four-fermion interactions. This ultimately explains the difference in the force laws. It is instructive to `derive' the $1/r^2$ law using dimensional analysis. First of all, since we are interested only in the contribution to the force due to the exchange of fermions between different points in spacetime, only the second order diagram in Figure \ref{loopdiag2} is relevant for us. We will work in Euclidean space and regularize the volume by considering a finite 4-dimensional box with edges of length $L$. In the end we will take the limit $L\rightarrow\infty$. Using that the fermion propagator in position space behaves like $\sim1/|x|^3$, its contribution to the vacuum energy (see \eqref{vacenergy}\footnote{In the case at hand we have $T=L$.}) takes the form
\begin{equation}
 E\sim\tilde{v}^2\frac{1}{L}\int d^4xd^4y\frac{1}{|x-y|^6}\theta(x)\theta(y)\,,
\end{equation}
where the source $\theta(x)$ is chosen according to \eqref{thetawall} and can be represented by $\theta(x)=\theta_1+(\theta_2-\theta_1)(\Theta(x_1-a)-\Theta(x_1-b))$ with $\Theta$ being the Heaviside step function. In order to avoid confusion we have changed notation compared to \eqref{thetawall} such that $x_1$ denotes the first component of the Euclidean four-vector $x$ corresponding to the direction orthogonal to the domain walls. By appropriately shifting the integration variable $y$ we can bring the integral into the form
\begin{equation}
 E\sim\tilde{v}^2\frac{1}{L}\int d^4xd^4y\frac{1}{|y|^6}f(x_1,y_1,b-a)\sim\tilde{v}^2L^2\int dx_1dy_1\frac{1}{|y|^3}f(x_1,y_1,b-a)\,
\end{equation}
for some function $f$, which demonstrates that the final result can only depend on $L$ and $b-a$. In general, there will be infinite contributions to the energy in the limit $L\rightarrow\infty$. A quantity that has a chance of being finite is the energy per unit domain wall area, $E/L^2$, whose finite part must be $\sim\tilde{v}^2/(b-a)$ for dimensional reasons. Upon differentiation this reproduces the $1/r^2$ behavior of the force density which was obtained in the exact analysis.

Let us now go back to our original question whether the NLO corrections to the force density found above spoil the validity of the 3-form effective description and the 1-instanton approximation. Naively, i.e.~ignoring the NLO force corrections, these approximations are valid in the energy regime $0\le\mu\le m$ which corresponds to distances $b-a$ with $m(b-a)\gg1$. As long as the ratio $\xi$ of the NLO contribution in \eqref{forcefermions} (without the mass renormalization term and using the approximation \eqref{forcecorrection}) to the leading contribution \eqref{forcepure} is much smaller than one this remains also true if we take the force corrections into account. We find
\begin{equation}
 \xi\lesssim\frac{m}{v}\cdot\frac{\text{e}^{-2m(b-a)}}{(m(b-a))^{3/2}}\cdot S^{2N}\text{e}^{-S}\cdot\frac{(2N)!}{N!}\text{e}^{-N} \,,
\end{equation}
where we used the formulae \eqref{kprime} and \eqref{kappa} for $K'$ and $\kappa$, respectively. For $\xi\ll1$ the 3-form description is still valid. Since we are in the regime $m/v\ll1$, $m(b-a)\gg1$ and $S\gg1$, the first three factors are much smaller than unity. However, the last term exhibits a factorial growth with $N$. While this may spoil the condition $\xi\ll1$ and hence the validity our 3-form EFT in principle, we can always avoid this by keeping $N$ sufficiently small.

Finally we are in a position to address the question, which was raised in the Introduction, whether there exists an emergent bosonic degree of freedom in Higgsed YM theory with massless fermions. If this would be the case, we expected either a constant contribution to the force between the domain walls for a massless boson or a purely exponential contribution in the case of a massive boson. Neither of this is the case for the force in (\ref{forcefermions}). Hence, we conclude that there is no sign for the presence of some emergent bosonic degree of freedom and the sub-leading corrections to the force are really due to the exchange of multiple fermions according to the 't Hooft interaction in (\ref{partfuncthooft}).  

There is, however, a loophole to our conclusion that no bosonic degree of freedom is present. In the general case of $N_\text{f}$ flavors the 't Hooft interaction is a $2N_\text{f}$-fermion interaction and we therefore have a Nambu-Jona-Lasinio type effective theory for the fermions \cite{Nambu:1961tp,Nambu:1961fr}. For such theories it has been shown that non-perturbatively generated masses for the fermions and bosonic bound states of fermions are present at large enough coupling $g$ \cite{Caldi:1977rc,Carlitz:1977fn}. In particular this is what is thought to be happening in QCD, leading to chiral symmetry breaking. However, we have limited our analysis to the small coupling regime, $g\ll1$, so we expect that this is not relevant in our case.

\subsection{Yang-Mills theory with fermions and Yukawa couplings}

\label{subsec::yukawa}

So far we have seen that Higgsed YM theory with light fermions, i.e.~lighter than the Higgs scale, can effectively be described by a 3-form theory,
\begin{equation}
 \mathcal{L}=\frac{1}{2\Lambda^4}(*F_4)^2+\theta(*F_4)\,, \label{3form}
\end{equation}
at scales below the fermion mass $m$ and with $\Lambda^4\sim(m/v)^{N_\text{f}}v^4$. This means in particular that this description completely breaks down in the limit $m\rightarrow 0$.

It is possible write down a model where fermions enforce the vanishing of $\Lambda$ {\it without} becoming massless themselves. To see this,\footnote{
We 
owe this idea and its simple model realization to Gia Dvali.
}
consider a Higgsed SU(2) gauge theory with gauge coupling $g$ and which is Higgsed by the vacuum expectation value of a scalar doublet $H$ with $\langle H^\text{T}\rangle=v(0,1)$. Furthermore, we add Weyl fermions in the following representations: two SU(2) doublets $Q_{1,2}$ and four singlets $\chi_i$ with $i\in\{1,2,3,4\}$. We generate masses via Yukawa couplings
\begin{equation}
\mathcal{L}_\text{Y}=y_1H^\dagger Q_1\chi_1+y_2\epsilon_{\alpha\beta}H^\alpha Q_1^\beta\chi_2+y_3H^\dagger Q_2\chi_3+y_4\epsilon_{\alpha\beta}H^\alpha Q_2^\beta\chi_4\,,
\end{equation}
where $\alpha,\beta$ are SU(2) indices. For simplicity we choose the Yukawa coupling constants $y_i=1$ such that the fermions obtain masses of the order of the Higgs scale $v$. Finally we add an explicit mass term
\begin{equation}
\mathcal{L}_M=M\epsilon_{\alpha\beta}Q_1^\alpha Q_2^\beta\,.
\end{equation}

It is crucial to note that whatever value $M$ has, the fermion masses will always be at least $v$ due to the Yukawa couplings. Consider a U(1) transformation according to which $Q_{1,2}\rightarrow\text{e}^{\text{i}\alpha}Q_{1,2}$ and $\chi_i\rightarrow\text{e}^{-\text{i}\alpha}\chi_i$. This U(1) is anomalous with respect to SU(2) and explicitly broken by the mass term $\mathcal{L}_M$. Thus the SU(2) $\theta$-parameter is physical and below the Higgs scale the theory is effectively described by a 3-form theory like \eqref{3form}. The cutoff of this effective theory is given by $m$.

Let us consider how the effective 3-form description is affected by the parameter $M$. For $M=0$ the anomalous U(1) can be used to rotate away the SU(2) $\theta$-parameter. It is hence unphysical. However, in the effective 3-form description $\theta$ is still physical which seems to be a contradiction. Furthermore, for all $M>0$ there is no reason for the 3-form description to break down. In particular its cutoff $v$ is independent of the value of $M$. So how can the two points $M=v$ and $M=0$ be smoothly connected to each other in the effective 3-form description?

A reasonable and simple answer is that the 3-form decouples in the limit $M\rightarrow0$. By this we mean that $\Lambda^2=\Lambda^2(M)$ such that $\Lambda^2(0)=0$. In this way the 3-form is unable to generate a potential for the $\theta$-parameter and makes it effectively unphysical. Hence consistency with the UV theory is restored. We expect that the coupling constant $\Lambda^2$ of the 3-form theory must be proportional to some positive power of $M$. Its role is analogous to that of the fermion masses in our original model without Yukawa couplings. In the following we will assume that this analogy can be taken literally and $\Lambda^2(M)$ is given by \eqref{lambdainstfermi1} with $m=M$. Note also that the decoupling of the 3-form is due to the change of a parameter of the UV theory and takes place without changing the degrees of freedom in the IR.

\section{Swampland constraints on axions and fermions}

\label{sec::swamp}

\subsection{Global symmetries and fermion operators}

\label{subsec::globalsym}

Axions have by definition a perturbative global shift symmetry. If this symmetry were exact also at the non-perturbative level, it would violate the quantum gravity censorship of global symmetries. We therefore expect this symmetry to be broken by non-perturbative effects in a consistent theory.

Indeed, this is realized if the axion couples to a (Higgsed) YM theory with $N_\text{f}$ massive gauged Dirac fermions. In the following we will refer to such a model as the light-fermion-scenario in contrast to the heavy-fermion-scenario explained in Subsection \ref{subsec::yukawa}. This terminology is supposed to stress the fact that the fermions in the model presented in Subsection \ref{subsec::yukawa} remain always massive due to the Yukawa couplings. The Lagrangian of the light-fermion-scenario reads 
\begin{equation}
\mathcal{L}=-\frac{1}{4g^2}\text{tr}(F_{\mu\nu}F^{\mu\nu})+\sum_{i=1}^{N_\text{f}}\overline{\psi}_i(\text{i}\slashed{D}-m_i)\psi_i-\frac{1}{2}f^2(\partial\phi)^2-\frac{\phi}{16\pi^2}\text{tr}(F_{\mu\nu}\tilde{F}^{\mu\nu})\,,\label{ym} \footnote{We have suppressed the Higgs sector here.} 
\end{equation}
where $\phi$ denotes the axion and $f$ its decay constant. Now the shift symmetry is non-perturbatively broken by instantons which induce an effective potential for the axion at low energies (below the Higgs scale).

If at least one of the fermions becomes massless the Lagrangian becomes invariant under the transformation $\psi_i\rightarrow\text{e}^{\text{i}\alpha\gamma_5}\psi_i$ and $\phi\rightarrow\phi-2\alpha$, where $\psi_i$ is the massless fermion. Since this symmetry contains a shift in the axion and is exact at the quantum level, we conclude that the axion potential must vanish in the presence of a massless fermion.\footnote{Strictly speaking we do not have an exact global symmetry here because of the chiral gravitational anomaly. However, this effect is severely suppressed, as we will discuss in the next section, and can be eliminated by adding an appropriate number of ungauged massless fermions.} This can be viewed as a simple symmetry argument for a technical result of the instanton calculus. However, as we will discuss later on, we expect such a theory to be constrained due to the quantum gravity censorship of global symmetries \cite{Abbott:1989jw,Coleman:1989zu,Kallosh:1995hi,Banks:2010zn}.

Let us now consider the heavy-fermion-scenario of Subsection \ref{subsec::yukawa}. It consists of a Higgsed SU(2) YM theory with two Weyl fermion doublets $Q_1,Q_2$ and four Weyl singlets $\chi_1,\chi_2,\chi_3,\chi_4$. These eight Weyl fermions are given a mass via the Higgs mechanism such that we end up with four massive Dirac fermions. Furthermore, we add an explicit mass term $M\epsilon_{\alpha\beta}Q_1^\alpha Q_2^\beta$. If we couple an axion to this theory, instantons will generate an effective axion potential. In the limit $M\rightarrow0$ the theory becomes invariant under an exact global symmetry with the transformation law $Q_{1,2}\rightarrow \text{e}^{\text{i}\alpha}Q_{1,2}$, $\chi_i\rightarrow\text{e}^{-\text{i}\alpha}\chi_i$ and $\phi\rightarrow\phi-2\alpha$. Similarly to the light-fermion-scenario we can conclude from this that the axion potential must vanish in the limit $M\rightarrow0$.\footnote{
As already noted in the Introduction, a closely related situation arises for an SU(2)$_L$ axion extending the Standard Model. In this case, the anomalous symmetry is U(1)$_{B+L}$ and its possible explicit breaking by higher-dimension operators has been used to argue for a ultra-light axion in~\cite{Nomura:2000yk}.
}
The existence of this global symmetry is again in conflict with quantum gravity expectations.

At first sight the two examples given above may look very similar. Here we would like to point out an important difference. In both theories the axion potential becomes zero in the limit of a vanishing mass parameter. While in the light-fermion-scenario this results in the presence of a truly massless fermion, in the heavy-fermion-scenario all fermions remain massive in the limit $M\rightarrow0$ due to the Yukawa couplings. In particular, in the former case the axion potential vanishes only at the expense of changing the IR degrees of freedom by introducing massless fermions while in the latter case those degrees of freedom are fixed for all values of $M$.

We have seen that both examples have a global symmetry which eventually allows for the presence of an exactly massless axion. One may argue that those two theories are simply incompatible with quantum gravity and hence reside in the swampland. However, a similar situation arises in $\mathcal{N}=2$ SUGRA which contains an axion with an exactly flat potential \cite{DAuria:1990qxt}. Nevertheless, we expect higher fermion interaction terms to break this symmetry explicitly and thereby make the theory consistent with quantum gravity again. Similarly, such additional fermion interactions could be used to make our two scenarios consistent with quantum gravity, too.

Motivated by this we conjecture that such additional fermion interactions are mandatory for fermions without a hard mass term in the presence of axions in order to prevent the existence of a global symmetry. This is a non-trivial statement and we find it interesting how the exclusion of a global axionic shift symmetry imposes constraints on fermion interactions. One could ask whether a minimal strength of these interactions can be inferred on general grounds and in the next two subsections we attempt to do so by conjecturing a constraint on axion potentials.

\subsection{(Too strong) a constraint on axions from the WGC for 3-forms}

Let us now bring back in the 3-form description of our two models. So far we have discussed the effective 3-form description of Higgsed YM theory without axions in Section \ref{sec::3eft}. However, according to \eqref{partfuncscalar1} the axion is easily accommodated by replacing the source $\theta$ by the axion field $\phi$ and adding a corresponding kinetic term for it. Then we immediately see that the axion mass $m_\phi^2$ is given by $m_\phi^2=\Lambda^4/f^2$. That means, whenever a global symmetry of the full UV theory forbids an effective axion potential, the effective 3-form description must decouple as $\Lambda^2=0$ (cf. Subsection \ref{subsec::yukawa}) is required for a vanishing potential. This is exactly what we observe in the case of a Higgsed YM theory in the limit of massless fermions and what we still expect to happen once Yukawa couplings have been introduced as in the heavy-fermion-scenario. More generally, the 3-form gauge coupling $\Lambda^2$ should always be proportional to a symmetry-breaking parameter of the UV theory.

The idea is now to apply the WGC to the effective 3-form description of YM theory and thereby derive a constraint on the 3-form gauge coupling and hence also on the axion potential. Let us start by stating the two versions of the WGC for 3-forms. The electric WGC for 3-forms requires the existence of domain walls which naturally couple to the 3-form and whose tension $T$ is bounded from above according to
\begin{equation}
 T\lesssim\Lambda^2M_\text{P}\,, \label{3formewgc}
\end{equation}
where $\Lambda^2$ is the 3-form gauge coupling \cite{ArkaniHamed:2006dz}. As long as the cutoff of the 3-form theory is below $(\Lambda^2M_\text{P})^{1/3}$ this bound has no consequences because the theory breaks down before the presence of such heavy domain walls is required by the WGC. On the other hand, if the cutoff is larger than $(\Lambda^2M_\text{P})^{1/3}$ and domain walls are not part of the theory, such a theory is forbidden by the WGC. The magnetic version of the WGC for 3-forms simply bounds the cutoff $\mu$ from above according to 
\begin{equation}
\mu\lesssim(\Lambda^2M_\text{P})^{1/3}\,, \label{3formmwgc}
\end{equation}
which is exactly the condition for the electric WGC to be satisfied without the presence of a light domain wall.

In order to apply the WGC to our effective 3-form theory of Higgsed YM theories we need to discuss the cutoff of this effective theory in some detail. In Section \ref{sec::3eft} we have found that the effective 3-form description of Higgsed YM theory is naively valid up to the Higgs scale $v$ while the presence of light fermions with masses $m\lesssim v$ reduce this cutoff down to $m$. If these fermions acquire their masses via order one Yukawa couplings, the cutoff stays at $v$. However, we have ignored a caveat in the corresponding argument which we would like to point out now. Recall from Section \ref{sec::3forms} that the energy density of the 3-form theory is given by $\epsilon=\Lambda^4\theta^2/2$. Furthermore, the effective 3-form description is only valid for $\theta\ll1$ and thus breaks down at energy densities corresponding to $\theta\sim1$, i.e.~$\epsilon\sim\Lambda^4$. Therefore the actual cutoff $\mu$ of the EFT is $\mu\sim\Lambda$. Using this new cutoff we find $(\Lambda^2M_\text{P})^{1/3}\gtrsim\Lambda\sim\mu$ which means that the theory always perfectly satisfies both the magnetic and electric WGC.

The last paragraph showed that the effective 3-form description of instantons breaks down at a scale set by the 3-form gauge coupling $\Lambda^2$. In particular this cutoff can be much lower than naively expected. Consider for example the heavy-fermion-scenario. In this theory the 3-form gauge coupling $\Lambda$ can be made arbitrarily small by choosing the parameter $M$ appropriately. In particular we can choose it such that we have a cutoff $\mu\sim\Lambda\ll v$, where $v$ is the Higgs scale as usual. On the other hand, there are no new degrees of freedom in the theory below the Higgs scale. Hence, there should be an effective theory that is valid in the energy range between $\mu\sim\Lambda$ and $v$ and contains the same degrees of freedom as the original effective 3-form theory.

In order to get rid of the constraint $\theta\ll1$ we would like to find a 3-form theory with action $S[F_4,\Lambda^2]$ such that it reproduces the full partition function \eqref{partfuncinstfull}, i.e.~
\begin{equation}
 \int\mathcal{D}A_3\exp\left(-S[F_4,\Lambda^2]-\text{i}\int\theta F_4\right)\propto\exp\left(\Lambda^4\int d^4x\cos\theta\right)\,,
\end{equation}
with $\Lambda^2$ satisfying \eqref{3formcouppure} or \eqref{lambdainstfermi1}, depending on whether we include fermions or not. Although we cannot determine the explicit form of $S[F_4,\Lambda^2]$, the above formula implicitly defines it and thereby also defines the effective 3-form theory we are looking for. In general  $S[F_4,\Lambda^2]$ can be a very complicated functional. For example, from the instanton calculus we expect $\exp(-S[F_4,\Lambda^2])$ to have non-trivial support only at $F_4$-configurations which are $\delta$-function-localized at certain points, each contributing one unit to $\int F_4\in\mathbb{Z}$.

Now we can use this improved effective 3-form theory and apply the WGC to it. According to \eqref{3formmwgc} the 3-form gauge coupling must obey $\Lambda^2\gtrsim(\mu/M_\text{P})\mu^2$. This turns out to be an extremely strong statement. To see this consider for example a simple Higgsed YM theory with Higgs scale given by $v$. As usual the corresponding effective 3-form theory has a gauge coupling $\Lambda^4=v^4\text{e}^{-S}$ and cutoff $\mu=v$. \eqref{3formmwgc} then implies
\begin{equation}
 v\lesssim\exp\left(-\frac{4\pi^2}{g^2}\right)M_\text{P}
\end{equation}
where $g$ is the gauge coupling constant of the YM theory evaluated at the scale $v$ and we have used the relation $S=8\pi^2/g^2$. This would imply that weakly coupled YM theories can only be spontaneously broken at exponentially low scales. Even though this is a valid result we think that it is too strong as there is naively no good reason why such a scenario should not be realizable in string theory.

Therefore we discard \eqref{3formmwgc} and conclude that the application of the WGC to the effective 3-form description of Higgsed YM theory leads to peculiar results. On the other hand, as we have discussed above, the WGC applied to the 3-form theory with the conservative estimate $\mu\sim\Lambda$ for the cutoff $\mu$ is satisfied. From this we conclude that, if the WGC for 3-forms has any regime of validity at all, it can only be applied to canonical 3-form theories with the standard action given by \eqref{actionpure}.

\subsection{A conjecture on axion potentials and implications for fermions}

Given our failed attempt to use the WGC for 3-forms to constrain axion potentials we now instead try to find a reasonable conjecture for a bound on the axion potential in the following. In general we expect the non-perturbative axion potential to be of the form
\begin{equation}
 V(\phi)=-V_0\text{e}^{-S}\cos\phi+\mathcal{O}\left(\text{e}^{-2S}\right)\,.
\end{equation}
We would like to find a lower bound on the amplitude of this potential. A first step into this direction is the WGC for axions which constrains the action $S$ according to $S\lesssim M_\text{P}/f$, i.e. $V_0\text{e}^{-S}\gtrsim V_0\text{e}^{-M_\text{P}/f}$ \cite{ArkaniHamed:2006dz}. However, as long as $V_0$ is completely free this does not provide a hard bound on the potential. Very naively one could conjecture that $V_0\text{e}^{-S}\gtrsim M_\text{P}^4\text{e}^{-M_\text{P}/f}$ but this, again, is too strong since the axion potential in $\mathcal{N}=2$ SUGRA vanishes exactly and therefore provides a counterexample in the landscape.

How can we reconcile a vanishing axion potential in SUSY moduli space and, at the same time, a lower bound on it? A possible answer is that the bound depends on the cutoff of the effective axion theory and vanishes for zero cutoff. We therefore propose the following general form of a bound on axion potentials:
\begin{equation}
 V_0\text{e}^{-S}\gtrsim\mu^\alpha M_\text{P}^{4-\alpha}\exp\left(-c\frac{M_\text{P}}{f}\right)\,. \label{axionpotcon}
\end{equation}
Here $\mu$ is the UV cutoff of the low energy theory that exclusively contains the axion while $\alpha>0$ and $c={\cal O}(1)$ parametrize our ignorance of the exact form of the bound.\footnote{In principle, $c$ is fixed by the precise form of the WGC for axions: $S\le cM_\text{P}/f$. But an exact analogue of extremal black holes, which normally define the bound, does not exist. Natural candidates might be axionic wormholes \cite{Giddings:1987cg,Hebecker:2018ofv}, suggesting the value $c=\pi\sqrt{6}/4$. Alternatively, if wormholes are absent, extremal gravitational instantons may define the bound~\cite{Heidenreich:2015nta}. This, however, may not be universal since different stringy models have different saxion couplings leading to slightly different values for $c$~\cite{Hebecker:2016dsw}.

Let us also sketch a supergravity example (\`a la KKLT) hinting towards $\alpha=2$. Consider a superfield $T=\tau+\text{i}\phi$ with shift-symmetric K\"ahler potential $K(T+\overline{T})$ and constant superpotential $W=W_0$. In principle both objects can receive additive instanton corrections of the form $\sim\text{e}^{-T}$. If only $K$ is corrected and if $|W_0|\ll1$, the leading contribution to the axion potential for $\phi$ is $|D_TW|^2\supset\sim\text{e}^{-\tau}m_{3/2}^2$, where $m_{3/2}$ is the gravitino mass. It appears plausible that $m_{3/2}$ is (of the order of) the mass of the lightest fermion and provides the cutoff $\mu$ relevant for the pure-axion theory. This suggests the bound $V\gtrsim M_\text{P}^2\mu^2 \text{e}^{-M_\text{P}/f}$ on axionic potentials. \label{fn}} 
The corresponding bound on the axion mass is
\begin{equation}
 m_\phi\gtrsim\frac{M_\text{P}\mu}{f}\left(\frac{M_\text{P}}{\mu}\right)^{1-\alpha/2}\exp\left(-c\frac{M_\text{P}}{2f}\right)\,. \label{axionmasscon}
\end{equation}
In order for this to be consistent with the fact that $\mu$ is the cutoff of the pure axion theory, this lower bound must lie below $\mu$. For $f\ll M_\text{P}$ this is generically true due to the exponential suppression by $\text{e}^{-cM_\text{P}/f}$. It is also interesting to consider the case $f\sim M_\text{P}$, even though this is not the regime we focused on. In this case $\text{e}^{-cM_\text{P}/(2f)}\sim\mathcal{O}(1)$ such that
\begin{equation}
 m_\phi\gtrsim\left(\frac{M_\text{P}}{\mu}\right)^{1-\alpha/2}\mu\,.
\end{equation}
This bound has a chance of being smaller than $\mu$ only if $\alpha\ge2$ which, interestingly, seems to be marginally satisfied in the supergravity example discussed in Footnote \ref{fn}.

Now let us discuss the implications of this bound for the two scenarios discussed so far. For simplicity we use $\alpha=4$ and set $c=1$ in the following. Let us start by considering the light-fermion-scenario with $N_\text{f}$ fermions of mass $m<v$ and an axion as defined by \eqref{ym}. The cutoff of the EFT of the axion is given by $\mu=m$. In order to satisfy \eqref{axionpotcon} the following inequality must hold
\begin{equation}
\left(\frac{m}{v}\right)^{N_\text{f}}\text{e}^{-S}\gtrsim\left(\frac{m}{v}\right)^4\exp\left(-\frac{M_\text{P}}{f}\right)\,.
\end{equation}
Here we have used \eqref{lambdainstfermiN} to determine $V_0$ in \eqref{axionpotcon}. Taking the WGC $S\lesssim M_\text{P}/f$ into account, this is trivially satisfied for any value of $m$ for $N_\text{f}\le4$ while $N_\text{f}>4$ implies the bound
\begin{equation}
 m\gtrsim\exp\left[-\frac{1}{N_\text{f}-4}\left(\frac{M_\text{P}}{f}-S\right)\right]>\exp\left(-\frac{M_\text{P}}{f}\right)v\,.
\end{equation}

Consequently the fermion mass is only restricted for $N_\text{f}>4$. For $N_\text{f}\le4$ an exactly flat axion potential is possible at the expense of massless fermions. In this case the degrees of freedom in the IR change and there exists no low energy theory that contains only the axion. As already discussed in Subsection \ref{subsec::globalsym}, such a theory has a global symmetry consisting of a shift in the axion and an anomalous U(1) rotation of fermions. We therefore expect additional fermion interactions to be present that break this symmetry. The constraint \eqref{axionmasscon} on the axion mass reads
\begin{equation}
  m_\phi\gtrsim\frac{m}{f}\exp\left(-\frac{M_\text{P}}{2f}\right)m\,.
\end{equation}

Next consider the heavy-fermion-scenario. To be more general consider the $N_\text{f}$-fold duplicated version of the model we have discussed so far, such that we have $N_\text{f}$ explicit mass terms with mass $M$. With $\mu=v$ \eqref{axionpotcon} implies
\begin{equation}
  M\gtrsim\exp\left(-\frac{1}{N_\text{f}}\left[\frac{M_\text{P}}{f}-S\right]\right)v\gtrsim\exp\left(-\frac{M_\text{P}}{f}\right)v
\end{equation}
which is very similar but stronger than what we have found in the last paragraph. Finally \eqref{axionmasscon} reads in this case
\begin{equation}
  m_\phi\gtrsim\frac{v}{f}\exp\left(-\frac{M_\text{P}}{2f}\right)v\,.
\end{equation}
This procedure could also be used to constrain other fermion operators that break a global symmetry which protects the axion potential.

\section{Gravitational instantons and fermion interactions}
\label{sec::grav}

Consider YM theory with $N_\text{f}$ massless Dirac fermions $\psi_i$, $1\le i\le N_\text{f}$, in the fundamental representation of the gauge group. This theory has an axial $\text{U}(1)_\text{A}$ symmetry on the classical level which is anomalously broken by instantons \cite{Vainshtein:1981wh}. As a result, the corresponding current $J^\mu_\text{A}$ of the symmetry is not conserved:
\begin{equation}
 \partial_\mu J_\text{A}^\mu=-\frac{N_\text{f}}{8\pi^2}\text{tr}(F_{\mu\nu}\tilde{F}^{\mu\nu})
 \label{anomaly}
\end{equation}
with $\tilde{F}_{\mu\nu}=(1/2)\epsilon_{\mu\nu\rho\sigma}F^{\rho\sigma}$ and
\begin{equation}
 J_\text{A}^\mu=\sum_{i=1}^N\overline{\psi}_i\gamma^\mu\gamma_5\psi_i\,.
\end{equation}
By recalling that instantons are topologically non-trivial field configurations, obeying
\begin{equation}
 \frac{1}{16\pi^2}\int d^4x\,\text{tr}(F_{\mu\nu}\tilde{F}^{\mu\nu})=n\in\mathbb{Z}\,,
\end{equation}
we can conclude that
\begin{align}
-2nN_\text{f}=-\frac{N_\text{f}}{8\pi^2}\int d^4x\,\text{tr}(F_{\mu\nu}\tilde{F}^{\mu\nu})=\int d^4x\partial_\mu J_\text{A}^\mu & =\int_{-\infty}^{+\infty}dt\frac{\partial}{\partial t}\int d^3xJ_\text{A}^0 \nonumber \\ & = Q_\text{A}(t=+\infty)-Q_\text{A}(t=-\infty)\,.
\end{align}

This shows that the axial charge $Q_\text{A}$ must change by $-2N_\text{f}$ along a single instanton event. Since $Q_\text{A}$ counts the number of right-handed minus the number of left-handed fermions, an instanton must convert $N_\text{f}$ right-handed fermions into $N_\text{f}$ left-handed fermions.\footnote{To conclude this we have to use the fact that the sum of right- and left-handed fermions is conserved.} Hence we conclude that instantons induce a $2N_\text{f}$-fermion interaction that explicitly breaks the axial $\text{U}(1)_\text{A}$ symmetry. These are the well-known 't Hooft interactions and they can be explicitly calculated. In particular, it turns out that the vectorial $\text{U}(1)_\text{V}$ and the chiral flavor symmetry $\text{SU}(N)_\text{V}\times\text{SU}(N)_\text{A}$ are left unbroken by the interaction. Simply by using this symmetry breaking pattern we can construct the flavor structure of the 't Hooft vertex. To do so we define the $\text{U}(1)_\text{V}$-invariant matrix $\Psi_{ij}=\overline{\psi}_{\text{R}i}\psi_{\text{L}j}$ with $\psi_\text{L/R}=P_\text{L/R}\psi$. With this quantity we can construct exactly one interaction that is invariant under the chiral flavor symmetry, namely $\det\Psi$. We can not fix the color index structure by this line of reasoning but it will not be relevant for us anyway.

Having discussed the YM case let us now turn to gravity. For this it is most convenient to use Weyl fermions instead of Dirac fermions. Hence let us consider $N_\text{f}$ massless Weyl fermions $\chi_i$ with $1\le i\le N_\text{f}$ which live on a manifold with metric $g_{\alpha\beta}$. In terms of these the current $J_5^\mu$ reads
\begin{equation}
J_\text{A}^\mu=-\sqrt{|\det g_{\alpha\beta}|}\sum_i\overline{\chi}_i\overline{\sigma}^\mu\chi_i
\end{equation}
and there exists a corresponding gravitational chiral anomaly \cite{Delbourgo:1972xb,Eguchi:1976db,Fujikawa:1980eg,AlvarezGaume:1983ig} given by
\begin{equation}
 \partial_\mu J_\text{A}^\mu=-\frac{N_\text{f}}{384\pi^2}R_{\mu\nu\rho\sigma}\tilde{R}^{\mu\nu\rho\sigma}\,, \label{anomalygravity}
\end{equation}
where $R_{\mu\nu\rho\sigma}$ is the Riemann tensor and $\tilde{R}_{\mu\nu\rho\sigma}=(1/2)\epsilon_{\mu\nu\alpha\beta}R^{\alpha\beta}_{\ \ \rho\sigma}$. There exist a number of different gravitational instantons for which the integral of the right hand side of \eqref{anomalygravity} is non-zero \cite{Eguchi:1980jx} and hence induces a change in the charge associated with the current $J_\text{A}^\mu$. Among them are, for example, the well-known Eguchi-Hanson instanton or the K3 manifold \cite{Eguchi:1978xp,Eguchi:1978gw}. In the following we would like to treat gravitational instantons as fluctuations of flat spacetime $\mathbb{R}^4$. This can be done by cutting a 4-dimensional ball out of $\mathbb{R}^4$ so that the boundary of the resulting hole is $S^3$ which then can be connected to the boundary of the gravitational instanton via a wormhole-like throat. Sections of this throat are topologically $S^3$. In the case of the Eguchi-Hanson instanton this procedure is not possible since the topology of its boundary is $S^3/\text{Z}_2$ and hence does not match the $S^3$ of the hole in $\mathbb{R}^4$.\footnote{See, however, \cite{tHooft:1988wxy,Holman:1992ah,Arunasalam:2018eaz} and \cite{Deser:1980kc} for a discussion of possible physical effects of Eguchi-Hanson instantons and of gravitational instantons in general, respectively.}

\begin{figure}[h]
\begin{center}
\includegraphics[width=0.6\textwidth]{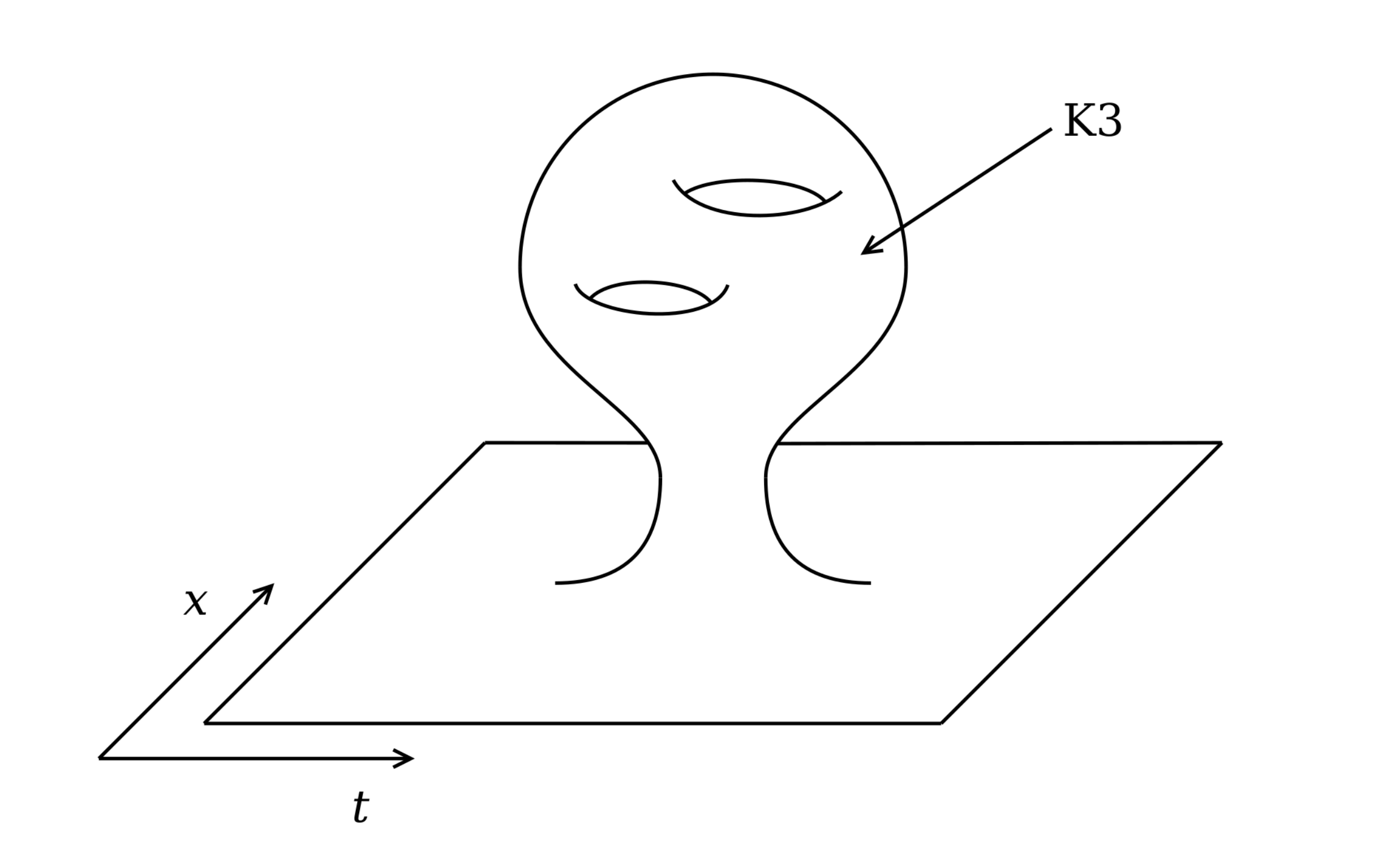}
\end{center}
\caption{A 2-dimensional illustration of a K3 instanton glued into flat spacetime.}
\label{instantonpic}
\end{figure}

A more promising candidate for a quantum fluctuation of $\mathbb{R}^4$ is the K3 manifold\cite{Hawking:1979zw,Hawking:1979pi}. Since it is Ricci-flat, it solves the vacuum Einstein equations and has a vanishing action. Furthermore, it is the only compact 4-dimensional manifold  with self-dual curvature such that the right hand side of (\ref{anomalygravity}) is non-zero \cite{Eguchi:1978gw,Hawking:1979zw,Hawking:1979zs,Eguchi:1980jx}, namely
\begin{equation}
 \frac{1}{48\pi^2}\int_\text{K3}d^4xR_{\mu\nu\rho\sigma}\tilde{R}^{\mu\nu\rho\sigma}=16\,. \label{k3top}
\end{equation}
As in the case of gauge instantons this result depends exclusively on the topology of K3 and is related to the Atiyah-Singer index theorem \cite{Eguchi:1980jx}. The topological nature of this relation will be important for us in the following. Since K3 has no boundary, we have to cut out a 4-dimensional ball in order to glue it into flat spacetime according to the general procedure described in the previous paragraph (see Fig.~\ref{instantonpic}). The resulting manifold is of course not a K3 anymore and has changed its topology. It also does not solve the Einstein equations anymore. Therefore one may be worried whether the topological relation (\ref{k3top}) still holds for the new manifold. In the following we argue why there is no problem.

Let us start with K3 that is glued onto an $S^4$ instead of $\mathbb{R}^4$ as described above. This manifold has no boundary and is topologically still a K3. Hence (\ref{k3top}) remains true. In the next step we split the anomaly integral into three parts, corresponding to the $S^4$, the original K3 contribution and the wormhole connecting them:
\begin{equation}
 \frac{1}{48\pi^2}\int d^4xR_{\mu\nu\rho\sigma}\tilde{R}^{\mu\nu\rho\sigma}=A_{S^4}+A_\text{K3}+A_\text{WH}\,. \label{gravtop}
\end{equation}
But the metrics on $S^4$ and on the wormhole have $R_{\mu\nu\rho\sigma}\tilde{R}^{\mu\nu\rho\sigma}=0$ locally and hence also $A_{S^4}=0=A_\text{WH}$. Now delete a point from the $S^4$ which gives $\mathbb{R}^4$ but certainly does not change the integral $A_{S^4}$. Hence we can conclude that a K3 glued into a flat region indeed gives a contribution to the anomaly integral. 

Finally, this allows us to conclude, similarly to the gauge instanton case, that a K3 fluctuation is, according to (\ref{anomalygravity}) and (\ref{k3top}), accompanied by a change of the axial charge $\Delta Q_\text{A}=-2N_\text{f}$. Hence K3 must induce an effective fermion interaction which is a product of $2N_\text{f}$ fermion fields. Since fermionic fields anti-commute and our theory contains exactly $2N_\text{f}$ of them (two components for each of the $N_\text{f}$ Weyl fermions) this information uniquely fixes the structure of this effective fermion interaction. It is simply the product of all fermion fields:
\begin{equation}
\mathcal{L}_\text{int}\propto\prod_i\chi_i^2=\prod_i(\chi_i)_1(\chi_i)_2\,, \label{K3int}
\end{equation}
where $(\chi_i)_1$ and $(\chi_i)_2$ denote the first and second component of the Weyl spinor $\chi_i$, respectively. The anti-commutativity of the fields implies that this product is totally antisymmetric in all of its indices, i.e.~in both flavor and spinor indices. Thus it is invariant under any subgroup of the full SU($N_\text{f}$) flavor symmetry as well as under any non-anomalous U(1) transformation of the fermions. This ensures in particular that the interaction is not in conflict with any pattern of consistent  gauge symmetries that could act on the fermion fields. 

Now we would like to estimate the strength of the effective interaction \eqref{K3int} by evaluating the contribution of K3 to the path integral of Euclidean quantum gravity with an insertion of this operator. To do so we need to integrate over all metrics of an asymptotically flat spacetime that contains one K3 fluctuation. Each such metric is suppressed by the action according to $\text{e}^{-S}$ with
\begin{equation}
 S=-\frac{M_\text{P}^2}{2}\int d^4x\sqrt{\det(g_{\mu\nu})}R[g_{\mu\nu}]
\end{equation}
and Ricci scalar $R[g_{\mu\nu}]$ which depends on the metric $g_{\mu\nu}$. While the asymptotic $\mathbb{R}^4$ and the K3 have vanishing curvature and hence do not contribute to the action, the wormhole part is non-trivial. Let $\rho$ be the typical size and curvature scale of the wormhole.\footnote{It may be interesting to analyze whether the controversial negative mode issue \cite{Coleman:1987rm,Rubakov:1996cn,Kim:1997dm,Kim:2003js,Alonso:2017avz,Hebecker:2018ofv,Hertog:2018kbz} of Giddings Strominger wormholes affects this wormhole region and hence K3 instantons as well.} Then, by dimensional analysis, its contribution to the action must be of the form\footnote{Note that wormholes are negatively curved which leads to the positive sign for the wormhole action.} $S_\text{WH}\sim\rho^2M_\text{P}^2$. The effective fermion interaction receives contributions from instantons of all sizes $\rho$ so that we have to integrate over it. Since we do not integrate out the fermions, no factor of the fermion mass can arise in front of the effective interaction. Therefore, we can use dimensional analysis to find
\begin{align}
 \mathcal{L}_\text{int} & \sim\int^{\infty}d\rho\,\rho^{3N_\text{f}-5}(\rho M_\text{P})^a\text{e}^{-\rho^2M_\text{P}^2}\prod_i\chi_i^2 \\
 & \sim\frac{1}{2}M_\text{P}^{4}\Gamma\left(\frac{3N_\text{f}+a-4}{2}\right)\frac{\prod_i\chi_i^2}{M_\text{P}^{3N_\text{f}}}=\mu^4\frac{\prod_i\chi_i^2}{\mu^{3N_\text{f}}}\,,
 \label{K3intfull}
\end{align}
where the last equality defines the typical scale $\mu$ of this interaction and $a$ parametrizes our ignorance about the instanton determinant. The integral over $\rho$ is dominated by a critical value $\rho_\text{c}^2=(3N_\text{f}+a-5)/(2M_\text{P}^2)$. If $N_\text{f}\sim\mathcal{O}(1)$ it is reasonable to assume $a\sim\mathcal{O}(1)$ as well since there is no large dimensionless number in the problem. Then one finds $\rho_\text{c}^{-1}\sim\mu\sim M_\text{P}$. By contrast, if $N_\text{f}\gg1$ and $a$ remains $\sim\mathcal{O}(1)$, one obtains $\rho_\text{c}^{-1}\sim \mu\sim M_\text{P}/\sqrt{N_\text{f}}$. In this case, the calculation remains controlled since the $\rho$ integral is dominated in the large-radius/weakly-curved regime. (However, $a$ might depend non-trivially on $N_\text{f}$, so we can not be certain about this.) For the Standard Model one would find $\mu\approx10^{17}\text{GeV}$, i.e.~an almost Planck-suppressed interaction.

Note that for $N_\text{f}=1$ \eqref{K3intfull} is a Majorana mass term. This raises the interesting question whether the K3 instantons generate fermion masses also for more than one flavor. A general mass term has the form
\begin{equation}
\sum_{i,j}\chi_iM_{ij}\chi_j \label{fermionmass}
\end{equation}
and can only be generated by \eqref{K3intfull} if it respects the symmetries of this interaction (this is analogous to the fact that chiral symmetry prevents fermion masses from being generated in perturbation theory). In the following we show that for $N_\text{f}>2$ we can always construct a non-anomalous U(1) transformation of the fermion fields which is a symmetry of \eqref{K3intfull} but not of \eqref{fermionmass} for any non-trivial mass matrix $M_{ij}$. Hence no masses are generated by \eqref{K3intfull} for $N_\text{f}>2$. To see this consider any non-vanishing term $\chi_iM_{ij}\chi_j$ in the sum \eqref{fermionmass}. Neither this term nor \eqref{K3intfull} is invariant under the anomalous U(1) transformation $\chi_i\rightarrow\text{e}^{\text{i}\alpha}\chi_i$. However, we can render it non-anomalous, and hence a symmetry of \eqref{K3intfull}, by letting a second field $\chi_k$ transform according to $\chi_k\rightarrow\text{e}^{-\text{i}\alpha}\chi_k$. For $N_\text{f}>2$ we can choose $k\neq j$ such that $\chi_iM_{ij}\chi_j$ is still not invariant under this transformation. Consequently, such mass terms can not be generated by \eqref{K3intfull}. So far this argument does not exclude a Dirac mass term for $N_\text{f}=2$. However, this possibility is also readily excluded by a symmetry argument. To do so note that a Dirac mass term of the form $\chi_1\chi_2$ breaks the discrete transformation $\chi_1\rightarrow- \chi_1$ while \eqref{K3intfull} remains invariant. To summarize, the K3 induced interaction does not generate masses for fermions except for $N_\text{f}=1$. \footnote{Euclidean wormholes may provide a different mechanism by which gravity breaks global symmetries, in particular inducing fermion interactions (see e.g.~\cite{Hawking:1988ae,Lyons:1989kh,Ellis:1989vk,Abbott:1989jw,Lee:1989zr}). However, wormholes also introduce deep conceptual problems, most notably with AdS/CFT~\cite{ArkaniHamed:2007js}, and may hence have to be excluded. This would enhance the role of K3 instantons in breaking global symmetries.}

In the following we show how the K3-induced interaction can nevertheless contribute to fermion mass generation under certain circumstances. Let us illustrate the idea with a simple example. Consider the case of $N_\text{f}=2$, i.e.~we have two Weyl fermions $\chi_1$ and $\chi_2$. Now let $\chi_1$ have a mass term $M\chi_1^2+\text{h.c.}$ while $\chi_2$ remains massless. Furthermore, the K3 induced interaction is a vertex with four external fermion lines, two associated with $\chi_1$ and two associated with $\chi_2$. Making use of the Feynman rules for Weyl fermions (see e.g. \cite{Dreiner:2008tw}) we see that the two $\chi_1$ lines can be connected to each other via the propagator $\sim M/(p^2-M^2)$ to give a diagram which generates a mass $m$ for $\chi_2$. Note that $M\ne0$ is crucial for this to give a non-zero contribution. Using \eqref{K3intfull} the full diagram can be estimated as
\begin{equation}
 m\sim\frac{1}{\mu^2}\int d^4p\,\frac{M}{p^2-M^2}\sim M\,\frac{\Lambda^2}{\mu^2}\,,
\end{equation}
where we have cut off the quadratically divergent momentum integral at the scale $\Lambda$. Since we are treating the K3 instantons as point-like and their typical size is $\rho_\text{c}^{-1}\sim\mu$, we have to impose $\Lambda\lesssim \mu$ for consistency. It is then natural to assume $\Lambda\sim\mu$, such that $m\sim M$.

This result can be readily generalized to the case of an arbitrary number $N_\text{f}$ of Weyl fermions of which all have a mass $M$ except for one. For this massless fermion a mass is generated by the K3 induced interaction if, again, all external lines of the vertex are connected to each other pairwise via the massive propagator. In this way a mass of order $m\sim(M\Lambda^2/\mu^3)^{N_\text{f}-1}\mu\sim M(M/\mu)^{N_\text{f}-2}$ for the originally massless fermion is generated. However, in the phenomenologically interesting case of low-scale masses $M\ll\mu$ we expect this to be miniscule. Indeed, for SM-like scales $M\sim1\ \text{GeV}$, $N_\text{f}\sim50$ and $\mu\sim M_\text{P}$ one finds $\ln(m/M_\text{P})\sim-1000$.

More generally it is possible to combine the K3 interactions also with higher-dimension operators to generate fermion masses.\footnote{In \cite{Nomura:2000yk} a similar mechanism has been used to generate an axion potential in the context of SU(2) gauge instantons.} This is even possible if all fermions are massless at tree level. Consider for example three massless Weyl fermions with K3 interaction $\chi_1^2\chi_2^2\chi_3^2$ and some additional interaction $\overline{\chi}_1^2\overline{\chi}_2^2$. This operator can be connected to the K3 vertex via four propagators to generate a mass for $\chi_3$. In contrast to the previous discussion here we connect external $\chi$-lines with external $\overline{\chi}$-lines which results in a propagator $\sim\overline{\sigma}^\mu p_\mu/(p^2-m^2)$ that does not vanish for zero mass, $m=0$. Note that the operator $\overline{\chi}_1^2\overline{\chi}_2^2$ and its complex conjugate alone do not generate a mass for any of the fermions. We want to emphasize that the effect of such combinations of K3 induced interactions and higher-dimension operators can in principle unsuppressed and hence large. It would therefore be interesting to analyze in more detail the effect of such operators in the SM. Experimental information, such as bounds on the proton lifetime, may then be used to constrain these operators. 

So far we have seen that K3 instantons alone in general do not generate fermion masses but can be combined with higher-dimension operators to do so. At this point we want to describe a mechanism by which one could possibly get rid of the need for such operators to obtain fermion masses. The idea is to have fermions charged under a U(1) gauge symmetry and to consider the effect of K3 instantons which have cycles carrying non-trivial U(1) flux.\footnote{The possibility of a K3 instantons with non-trivial U(1) flux has been noted in \cite{Deser:1980kc}} In this case the chiral gravitational anomaly \eqref{anomalygravity} would obtain corrections from the U(1) flux on the K3 according to
\begin{equation}
 \partial_\mu J_\text{A}^\mu=-\frac{N_\text{f}}{384\pi^2}R_{\mu\nu\rho\sigma}\tilde{R}^{\mu\nu\rho\sigma}-\frac{N_\text{f}}{32\pi^2}B_{\mu\nu}\tilde{B}^{\mu\nu}\,,
\end{equation}
where $B_{\mu\nu}$ denotes the field strength of the U(1) gauge symmetry and for simplicity we have assumed the charge of all fermions to have absolute value one. Upon integrating this equation over spacetime the second term on the right hand side will be proportional to the U(1) flux squared. The latter can be chosen to make the change in the charge associated with $J_\text{A}^\mu$ along such a K3 instanton small. This would then reduce the number of fermions which occur in the induced fermion interaction. If this number could be chosen to be two, this interaction would correspond to simple mass terms.

Furthermore, it would be interesting to determine whether these interactions may contribute to effective potentials of axions. If this axion couples to gravity via the topological term \eqref{gravtop}, this does not seem to be the case since a shift in the axion could be undone by an appropriate anomalous fermion transformation. However, a detailed discussion is needed to answer this question properly.

\section{Conclusions}
\label{sec::conc}

In the first part of this paper we have studied the effective 3-form description of instantons. To do so we coupled both theories to an external source $\theta$ and calculated the respective partition function and forces on domain walls which we modeled by a spatially varying $\theta$. While this calculation can be done exactly in the 3-form theory, for the gauge instantons one needs to employ the dilute gas approximation. This restricts the range of applicability to weakly coupled Higgsed YM theories. We found that the partition functions and forces agree for small values of the external source $\theta$ and an appropriately chosen 3-form gauge coupling constant $\Lambda^2$. This shows that 3-form theories indeed are EFTs of Higgsed YM theories for small $\theta$. We expect this correspondence to hold also for gravitational instantons. It would be interesting to see how the 3-form description can be improved such that the restriction to small $\theta$ can be relaxed.

With the same method we analyzed the effect of gauged fermions on the effective 3-form description. It turned out that they simply alter the expression for the gauge coupling $\Lambda^2$ of the effective 3-form theory by a factor proportional to their mass. This implies that massless fermions decouple the effective 3-form theory which is consistent with the fact that they completely suppress isolated gauge instantons. Recently, it has been argued that the effective 3-form description of YM theory with massless fermions could potentially contain a massless bosonic degree of freedom \cite{Dvali:2013cpa,Dvali:2016uhn,Dvali:2017mpy}. While this is the case in a confining theory like QCD with an $\eta'$ pseudoscalar, we are not able to find evidence for this in the case of a Higgsed YM theory.

After having discussed the effective 3-form description of instantons, we considered axionic shift symmetries in the second part of the paper. Ultimately, we expect such global symmetries to be broken due to quantum gravitational effects. Our intuition is then that this should manifest itself in terms of a non-vanishing potential and mass for the axion. Now, an interesting question is whether there is a quantitative bound on how small axion masses can be.

For YM theories coupled to fermions and an axion, the massless axion is protected by an exact global symmetry that involves a shift in the axion and an anomalous U(1) transformation of the fermions. This symmetry is generically broken by fermion operators which explicitly break the anomalous U(1) rotation of the fermions. Since quantum gravity censors global symmetries, it requires the presence of such operators, thereby also generating a mass for the axion. Conversely, if one has some a priori knowledge about a lower bound on axion masses, the coefficients of the relevant fermion operators can be constrained from below. 

However, we can not simply propose a general lower bound on axion masses in terms of $f$ and $M_\text{P}$. The reason is that $\mathcal{N}=2$ SUSY compactifications provide examples with exactly massless axions in the landscape. It is then natural to expect that any lower bound $\tilde{m}_\phi$ on axion masses must depend on the cutoff, $\tilde{m}_\phi=\tilde{m}_\phi(\mu)$. Here $\mu$ is the scale below which the effective theory contains exclusively the axion. If the $\mu$-dependence of the axion mass bound is such that $\tilde{m}_\phi(\mu=0)=0$, then consistency with $\mathcal{N}=2$ SUGRA is maintained. Indeed, these theories always have further massless degrees of freedom, such that the cutoff $\mu$ is zero.

Based on simplicity, the examples we have considered so far, and the WGC for axions we then propose the following bound on axion masses $m_\phi$:
\begin{equation}
m_\phi\gtrsim\frac{M_\text{P}^2}{f}\left(\frac{\mu}{M_\text{P}}\right)^{\alpha/2}\exp\left(-c\frac{M_\text{P}}{2f}\right)\,.
\end{equation}
Here $f$ is the axion decay constant, $\mu$ is the cutoff of the low-energy axion theory, and $\alpha>0$ is an unknown parameter. Since in YM theory instantons and fermions determine the non-perturbatively generated axion mass, one may then use this bound to constrain fermion masses $m$. The precise results are somewhat model dependent, but the structure is of the type
\begin{equation}
m\gtrsim\exp\left(-c\frac{M_\text{P}}{f}\right)v\,,
\end{equation}
where $v$ denotes the Higgs scale of the YM theory. 

There are at least two promising directions to make progress with this conjecture in the future. First of all, it is important to test it in stringy constructions. In particular, it would be interesting to understand more precisely how $\mathcal{N}=2$ SUSY or weakly broken ${\cal N}=1$ SUSY protect massless and very light axions respectively. Second, one can take our bound for granted and explore possible phenomenological implications. Especially constraints on fermion operators which break axionic shift symmetries could be studied in a variety of models.

Finally, in the last section we discussed the possibility of fermion operators which are generated by gravitational instantons. These operators are the analogues of the so-called 't Hooft interactions which are generated by gauge instantons. We argued that K3 instantons seem to be the only gravitational instantons capable of generating those interactions which, in a theory containing $N_\text{f}$ Weyl fermions $\chi_i$, take the form $\prod_i\chi_i^2$. A rough estimate of the strength of this operator showed that it is severely suppressed and not relevant for phenomenology. However, if combined with other higher-dimension fermion operators, it can give unsuppressed contributions to fermion masses. We also note that further interactions with different structure are possible if the fermions are charged under a U(1) gauge symmetry and K3 instantons with non-trivial U(1) flux are considered. In particular, this may allow for fermion mass terms which are directly induced by such gravitational instantons.

\subsection*{Acknowledgments}

We are very grateful to Gia Dvali for collaboration during part of this project and for many extremely useful discussions. We would also like to thank Paolo Di Vecchia, Archil Kobakhidze and Pablo Soler for helpful discussions. This work is supported by Deutsche Forschungsgemeinschaft (DFG) under Germany's Excellence Strategy EXC-2181/1 - 390900948 (the Heidelberg STRUCTURES Excellence Cluster).

\appendix

\section{3-form gauge theory}

\label{app::a}

\subsection{Pure 3-form gauge theory}

\label{app::pure}

The free theory of a 3-form gauge potential $A_3$ is defined by the Euclidean action
\begin{equation}
 S_\text{E}[A_3,\theta]=\int_{M_4}\frac{1}{2\Lambda^4}F_4\wedge*F_4-\text{i}\int_{M_4}\theta F_4\,, \label{puref4action}
\end{equation}
where $F_4=\text{d}A_3$ is the field strength associated to $A_3$ and $\theta$ is an external source. $\Lambda^2$ corresponds to the coupling constant and $M_4$ is the 4-dimensional Riemannian manifold on which the gauge theory lives. In the following we take $M_4=S^1\times M_3$ with $M_3$ having finite volume and no boundary unless otherwise stated. The corresponding (thermal) partition function is
\begin{equation}
 Z[\theta]=\int\mathcal{D}A_3\text{e}^{-S_\text{E}[A_3,\theta]}\,.
\end{equation}
We have normalized $F_4$ such that the Dirac quantization condition reads $\int_{M_4}F_4=n\in\mathbb{Z}$. Making use of this quantization condition we can rewrite the partition function as
\begin{equation}
 Z[\theta]=\int\mathcal{D}(*F_4)\sum_n\delta\left(\int_{M_4}F_4-n\right)\text{e}^{-S_\text{E}[F_4,\theta]}\,,
\end{equation}
where now we view $F_4$ as an independent integration variable and treat the action as a functional of $F_4$. After rewriting $\delta(\int_{M_4}F_4-n)=\int d\chi/(2\pi)\exp(\text{i}\chi(\int_{M_4}F_4-n))$, performing the $F_4$-integral and using the identity $\sum_n\exp(\text{i}\chi n)=2\pi\sum_n\delta(\chi-2\pi n)$ we find for the partition function
\begin{equation}
 Z[\theta]=C \sum_n\exp\left(-\frac{\Lambda^4}{2}\int_{M_4}(\theta+2\pi n)^2*1\right)\,,
\end{equation}
with $C$ being a possibly infinite constant.

In order to understand this theory physically let us consider the partition function $Z[\theta]$ in the limit of constant $\theta$. Then
\begin{equation}
  Z[\theta]=C \sum_n\exp\left(-\frac{\Lambda^4\beta V}{2}(\theta+2\pi n)^2\right)\,,
\end{equation}
where $\beta$ is the circumference of $S^1$ and $V$ denotes the volume of $M_3$. This is the partition function of a theory with infinitely many orthogonal energy eigenstates labeled by all integers $n$ and with energy given by $E_n=\Lambda^4V/2(\theta+2\pi n)^2$. From the form of $Z[\theta]$ it is clear that the theory is invariant under the shift $\theta\rightarrow\theta+2\pi$. Hence it is sufficient to consider only $\theta\in[-\pi,\pi)$.\footnote{Note that for $\theta=-\pi$ there are two degenerate energy eigenstates. We will ignore this subtlety in the following.} For this choice the vacuum energy is given by $E_0$. For $V\rightarrow\infty$ only the vacuum state remains while all other energy eigenstates disappear due to their exponential suppression relative to the vacuum. In the following we will only keep the vacuum state as we are primarily interested in the limit $V\rightarrow\infty$.

The partition function for infinite volume and arbitrary $\theta(x)$ reads
\begin{equation}
 Z[\theta]=C\exp\left(-\frac{\Lambda^4}{2}\int_{M_4}\theta^2*1\right)\,. \label{partfuncfreef4app}
\end{equation}
From this we easily read off the energy density
\begin{equation}
 \epsilon=\frac{\Lambda^4}{2}\theta^2 \label{endenf4}
\end{equation}
and calculate the vacuum expectation value of $*F_4$:
\begin{equation}
  \langle *F_4\rangle=\frac{1}{\text{i}Z[\theta]}\frac{\delta Z[\theta]}{\delta\theta}=\text{i}\Lambda^4\theta\,.
\end{equation}
We also find for the correlator of $*F_4$
\begin{equation}
 \langle *F_4(x)*F_4(0)\rangle=\Lambda^4\delta(x)\,.
\end{equation}
The appearance of the $\delta$-function in the correlator and the fact that the vacuum expectation value exactly follows the external source shows that the field strength is a purely local object that does not propagate any degree of freedom through spacetime.

So far we have seen that the 3-form theory in the large volume limit $V\rightarrow\infty$ has only one energy eigenstate, the vacuum, and lacks any propagating degrees of freedom. Therefore, one may be tempted to conclude that the theory does not contain any dynamics. This is not true as we will explain now. As is well known, the gauge potential $A_3$ naturally couples to the world-sheet (WS) of a domain wall via $\int_\text{WS}A_3$. Alternatively we may write this as $\int_{M_4}A_3\wedge J_1$ where $J_1$ is the conserved current of the domain wall. After integration by parts we see that the source term $\int_{M_4}\theta F_4$ is exactly of this form with $J_1=\text{d}\theta$. In the following we will choose an appropriate $\theta(x)$ that describes two domain walls and calculate the force that acts on them. It turns out that this force is not zero and therefore the theory is not trivial.

For simplicity we will choose $M_3$ to be $\mathbb{R}^3$ with coordinates $(x,y,z)$. We would like to describe two domain walls defined by $x=a$ and $x=b$. Assuming $a<b$ this is realized by the choice
\begin{equation}
 \theta(x)=
 \begin{cases}
  \theta_1 & \text{for } x\le a \\
  \theta_2 & \text{for } a < x < b \\
  \theta_1 & \text{for } b\le x
 \end{cases} \label{thetawall}
\end{equation}
with $\theta_1,\theta_2\in[-\pi,\pi)$. The force on the domain wall is simply the negative derivative of the energy associated to this configuration with respect to the position of the domain wall. However, this force is going to be infinite due to the domain walls being infinitely extended. Hence, the proper quantity to determine is the force per area. To do so we consider an infinite cylinder with a base area $A$ that is parallel to the domain walls. Now we calculate the change $\Delta E$ in the energy residing in the cylinder due to a small change $\Delta a>0$ in the position of the first domain wall. Using (\ref{endenf4}) we find  $\Delta E=A\Delta a(\Lambda^2/2)(\theta_1^2-\theta_2^2)$. Since this expression is linear in $A$ and $\Delta a$ we can immediately read off the force per area acting on the domain walls at $a$ and $b$, respectively,
\begin{equation}
 f^{(a)}=-f^{(b)}=\frac{\Lambda^4}{2}(\theta_2^2-\theta_1^2)\,. \label{forcepureapp}
\end{equation}

Note that the forces do not depend on the distance between the domain walls. The force on one of the walls remains non-zero even if we push the other domain wall out to infinity. The situation is somewhat analogous to that with a charged membrane positioned orthogonally to a homogeneous electric field.

\subsection{3-form gauge theory coupled to a scalar field}

Now we extend the 3-form theory by introducing a scalar field $\phi$ with mass $m$ that couples to $F_4$ according to the new action
\begin{equation}
 S_\text{E}[A_3,\phi,\theta]=\int_{M_4}\frac{1}{2\Lambda^4}F_4\wedge*F_4+\int_{M_4}\frac{f^2}{2}\text{d}\phi\wedge*\text{d}\phi+\int_{M_4}\frac{1}{2}m^2f^2\phi^2*1-\text{i}\int_{M_4}(\theta+\phi) F_4\,, \label{actionf4scalarapp}
\end{equation}
where $f$ determines the normalization of $\phi$. This theory is special for $m=0$ since in that case it is dual to a 2-form theory that is gauged by $A_3$. The dual action reads
\begin{equation}
 \tilde{S}_\text{E}[A_3,B_2]=\int_{M_4}\frac{1}{2\Lambda^4}F_4\wedge*F_4+\frac{1}{2f^2}\int_{M_4}(\text{d}B_2-A_3)\wedge*(\text{d}B_2-A_3) \label{actionf4b2}
\end{equation}
which can be easily checked by dualization under the path integral. This action is invariant under the simultaneous transformations $B_2\rightarrow B_2+\Omega_2$ and $A_3\rightarrow A_3+\text{d}\Omega_2$ for an arbitrary 2-form $\Omega_2$. It realizes the St\"uckelberg mechanism for $A_3$, i.e.~the gauge symmetry is spontaneously broken by the vacuum $B_2=\text{const.}$ such that only a massive $A_3$ is left. In the following we want to argue that on the $\phi$-side of the duality this symmetry breaking      can be possibly understood as an effect of the (quantum) dynamics of the massless $\phi$.

Let us make this argument for a more familiar example. Consider the following action:
\begin{equation}
 \tilde{S}_\text{E}[A_1,\varphi]=\int\frac{1}{2e^2}F_2\wedge *F_2+\int \frac{v^2}{2}(\text{d}\varphi-A_1)\wedge *(\text{d}\varphi-A_1)\,. \label{stueckelberg}
\end{equation}
This action realizes the St\"uckelberg mechanism for a 1-form gauge potential $A_1$. If we embedded this theory in a Higgs theory, $v$ would be the vacuum expectation value of the Higgs field. Hence we expect the gauge symmetry to be restored in the vacuum for $v=0$ which indeed is the case as is clear by inspection of the action.

Next let us have a look at the dual action which reads 
\begin{equation}
 S_\text{E}[A_1,B_2]= \int\frac{1}{2e^2}F_2\wedge *F_2+\int\frac{1}{2v^2}\text{d}B_2\wedge *\text{d}B_2-\text{i}\int B_2\wedge F_2\,. \label{stueckelbergdual}
\end{equation}
Since $B_2$ does not transform under the gauge symmetry of $A_1$ its classical vacuum configuration $B_2=\text{const.}$ does not break it. Let us inspect the case $v=0$ for which the spontaneous symmetry breaking is turned off. In this case the dynamics of the field $B_2$ is frozen and it effectively acts as a source for $F_2$. This observation suggests that the dynamics of $B_2$ is ultimately responsible for the spontaneous symmetry breaking. Note also that the duality of \eqref{stueckelberg} and \eqref{stueckelbergdual} breaks down when $B_2$ is massive. Hence this property of $B_2$ seems to be crucial for the dynamics behind the spontaneous symmetry breaking. All of these observations carry over to the theories defined by (\ref{actionf4scalarapp}) and (\ref{actionf4b2}).

Let us go back to the generic case with arbitrary $m$ and calculate the partition function of the theory. The $A_3$-integration can be carried out as before which leads to
\begin{equation}
 Z[\theta]=C\int\mathcal{D}\phi\exp\left(-\int_{M_4}\frac{\Lambda^4}{2}(\phi+\theta)^2*1-\int_{M_4}\frac{f^2}{2}\text{d}\phi\wedge *\text{d}\phi-\int_{M_4}\frac{1}{2}m^2f^2\phi^2*1\right)\,.
\end{equation}
This path integral is Gaussian in $\phi$ and we can therefore simply use the classical equation of motion,
\begin{equation}
\Box\phi=M^2\phi+(M^2-m^2)\theta \label{eomphi}
\end{equation}
with $M^2=m^2+\Lambda^4/f^2$, to find the formal result
\begin{equation}
 Z[\theta]=C'\exp\left(-\frac{\Lambda^4}{2}\int_{M_4}\theta(x)\frac{\Box-m^2}{\Box-M^2}\theta(x)*1\right)\,. \label{partfuncf4scalar}
\end{equation}
For $f\rightarrow\infty$, i.e.~$M\rightarrow m$, this reduces, up to constant factors, to (\ref{partfuncfreef4app}) as it should be. The vacuum expectation value and correlator of $*F_4$ are now calculated to be
\begin{equation}
 \langle *F_4\rangle=\text{i}\Lambda^4\frac{\Box-m^2}{\Box-M^2}\theta \label{vevf42}
\end{equation}
and
\begin{equation}
 \langle *F_4(x)*F_4(0)\rangle=\Lambda^4\frac{\Box-m^2}{\Box-M^2}\delta(x)\,.
\end{equation}
From the pole structure of the correlator we infer the presence of a massive degree of freedom with mass $M$ which continues to exist even for $m=0$. In fact this is not surprising as we have seen that for $m=0$ a St\"uckelberg mechanism is at work in the dual description.

Instead of using the formal expression (\ref{partfuncf4scalar}) to determine the force on domain walls we explicitly use a solution to the equation of motion with $\theta$ as defined in (\ref{thetawall}). This solution can be written as
\begin{equation}
\phi(x)= \left(1-\frac{m^2}{M^2}\right)
  \times\begin{cases}
  \frac{\theta_1-\theta_2}{2}(\text{e}^{M(x-a)}-\text{e}^{M(x-b)})-\theta_1 & \text{for }x\le a \\
  -\frac{\theta_1-\theta_2}{2}(\text{e}^{M(x-b)}+\text{e}^{-M(x-a)})-\theta_2 & \text{for } a < x < b \\
  \frac{\theta_1-\theta_2}{2}(\text{e}^{-M(x-b)}-\text{e}^{-M(x-a)})-\theta_1 & \text{for } b\le x
 \end{cases} 
 \,. \label{classolphi}
\end{equation}
Matching the solutions in the different regimes to each other at the boundary and demanding $\phi$ to be constant at $x\rightarrow\pm\infty$ fixes all six integration constants uniquely. Upon using the equation of motion (\ref{eomphi}) the action in (\ref{partfuncf4scalar}) can be rewritten as
\begin{equation}
 \int_{M_4}\frac{\Lambda^4}{2}\theta(x)(\phi(x)+\theta(x))*1\,.
\end{equation}
The integrand of this action is the energy density in the presence of the two domain walls. In order to appreciate its structure it is helpful to explicitly calculate it:
\begin{equation}
 \epsilon(x)=\frac{\Lambda^4}{2}
  \times\begin{cases}
  \frac{m^2}{M^2}\theta_1^2+\left(1-\frac{m^2}{M^2}\right)\theta_1\frac{\theta_1-\theta_2}{2}(\text{e}^{M(x-a)}-\text{e}^{M(x-b)}) & \text{for }x\le a \\
  \frac{m^2}{M^2}\theta_2^2-\left(1-\frac{m^2}{M^2}\right)\theta_2\frac{\theta_1-\theta_2}{2}(\text{e}^{M(x-b)}+\text{e}^{-M(x-a)}) & \text{for } a < x < b \\
  \frac{m^2}{M^2}\theta_1^2+\left(1-\frac{m^2}{M^2}\right)\theta_1\frac{\theta_1-\theta_2}{2}(\text{e}^{-M(x-b)}-\text{e}^{-M(x-a)}) & \text{for } b\le x
 \end{cases} 
 \,. \label{endenf4scalar}
\end{equation}
We clearly see that the first term equals the energy density (\ref{endenf4}) of the pure 3-form theory corrected by a factor $m^2/M^2$. The effect of this part of the energy density on the force per area is hence exactly as we have calculated in (\ref{forcepureapp}) but with the additional factor $m^2/M^2$. Now consider the second term in (\ref{endenf4scalar}). We would like to repeat the computation of the change in energy within a given cylinder as we have done in Subsection \ref{app::pure}. However, this time the energy density changes at arbitrarily large distances from the domain walls if we move them around. Hence, we have to use an infinitely extended cylinder. The total energy within such a cylinder with base area $A$, ignoring the first term in (\ref{endenf4scalar}) we have discussed already, is
\begin{equation}
 E=A\frac{\Lambda^4}{2}\left(1-\frac{m^2}{M^2}\right)\frac{1}{M}(\theta_1-\theta_2)^2(1-\text{e}^{-M(b-a)})\,.
\end{equation}
Taking the negative derivative with respect to $a$, dividing by $A$ and combing with the contribution from the first term in (\ref{endenf4scalar}) gives for the total force density
\begin{equation}
  f^{(a)}=-f^{(b)}=\frac{\Lambda^4}{2}\left[\frac{m^2}{M^2}(\theta_2^2-\theta_1^2)+\left(1-\frac{m^2}{M^2}\right)(\theta_2-\theta_1)^2\text{e}^{-M(b-a)}\right]\ \label{forcescalara}
\end{equation}

Let us compare this result with (\ref{forcepureapp}). We have again a constant contribution in (\ref{forcescalar}) which is suppressed by the factor $(m/M)^2$ compared to (\ref{forcepureapp}) and a new second term that exponentially falls off with the distance between the domain walls. Note that this exponential fall-off is exactly what we could have anticipated from the presence of a massive degree of freedom with mass $M$. While the first contribution to the force is due to the interaction of the domain walls with the background field strength, the second exponential term represents an interaction between the two domain walls due to a massive scalar field. In the limit $f\rightarrow\infty$, i.e.~in the decoupling limit of $\phi$, (\ref{forcescalara}) reduces to (\ref{forcepureapp}) as it should be. For $m=0$ the constant part of the force disappears while the second essentially remains unaffected. This can be intuitively understood by observing from (\ref{vevf42}) that the background field strength vanishes for $m=0$ and constant $\theta$. Hence there is no field strength the domain walls can interact with anymore and the corresponding force becomes zero. On the other hand, as already explained above, even though $m=0$ there is a massive scalar present which is why the second contribution to the force remains. 

\section{Instantons in Yang-Mills theory}

\label{app::b}

\subsection{Review of the instanton calculus}

\label{app::b1}

In this appendix we collect some well known results about gauge instantons. A good reference is for example \cite{Vainshtein:1981wh} but see also \cite{Bianchi:2007ft,Vandoren:2008xg,Blumenhagen:2009qh}. An SU($N$) gauge theory is described by the Euclidean action
\begin{equation}
 S_\text{E}[A_1,\theta]=\int\frac{1}{2g^2}\text{tr}(F_2\wedge*F_2)-\frac{\text{i}\theta}{8\pi^2}\text{tr}(F_2\wedge F_2)\,, \label{instactionapp}
\end{equation}
where $A_1$ is the Lie-algebra-valued gauge potential and $F_2=\text{d}A_1$. $g$ denotes the gauge coupling and $\theta$ can in principle be an external source that depends on space. Here we assume the topology of space to be simply $\mathbb{R}^4$. An instanton corresponds to a topologically non-trivial field configuration $A_1^\text{I}$ which minimizes the action and has the properties
\begin{equation}
 S_\text{E}[A_1^\text{I},\theta]=\frac{8\pi^2}{g^2}-\text{i}\theta\,.
\end{equation}
The instanton configuration $A_1^\text{I}$ has $4N$ moduli, four for the instanton location, one for its size and the rest for the orientation in group space. The contribution of an instanton with a given size $\rho$ and location $x$ to the partition function reads
\begin{equation}
\frac{d^4xd\rho}{\rho^5}C(N)f(g(\rho),N)\text{e}^{\text{i}\theta}\,, \label{instcont}
\end{equation}
where
\begin{equation}
 f(g(\rho),N)=\left(\frac{8\pi^2}{g^2(\rho)}\right)^{2N}\exp\left(-\frac{8\pi^2}{g^2(\rho)}\right)
\end{equation}
and
\begin{equation}
\frac{8\pi^2}{g^2(\rho)}=\frac{8\pi^2}{g^2(\rho_0)}-\frac{11}{3}N\ln\left(\frac{\rho}{\rho_0}\right)+\mathcal{O}\left[g^2(\rho_0)\ln\left(\frac{\rho}{\rho_0}\right)\right] \label{runningcoupling}
\end{equation}
takes into account the running of the coupling with the instanton size $\rho$. $\rho_0$ is an arbitrary reference scale. Furthermore, we have
\begin{equation}
 C(N)=\frac{C_1}{(N-1)!(N-2)!}\text{e}^{-C_2N}
\end{equation}
with $C_1$ and $C_2$ being order one numerical constants. Besides the instanton there is also an anti-instanton configuration $A_1^{\overline{\text{I}}}$ with
\begin{equation}
 S_\text{E}[A_1^{\overline{\text{I}}},\theta]=\frac{8\pi^2}{g^2}+\text{i}\theta
\end{equation}
and the corresponding contribution to the partition function is
\begin{equation}
\frac{d^4xd\rho}{\rho^5}C(N)f(g(\rho),N)\text{e}^{-\text{i}\theta}\,.
 \end{equation}
In the following we will use the abbreviation $S=8\pi^2/g^2(\rho_0)$.
 
Next we would like to determine the full contribution of instantons to the partition function. This can be done in the dilute gas approximation in which all instantons are considered point-like. Such an approximation is only valid if the density of instantons in space is small compared to their maximal size, i.e.~if there is no overlap between them. However, in principle we have to integrate \eqref{instcont} over all $\rho$ and hence take into account instantons of all sizes. In fact the contribution of large instantons, which are problematic for the dilute gas approximation, diverges. Indeed, inserting \eqref{runningcoupling} into \eqref{instcont} reveals that the integrand of the $\rho$-integration is given by $\rho^{11N/3-5}$. The exponent is positive for any $N\ge2$ which renders the integral IR divergent. Hence, the dilute gas approximation is not applicable in a pure non-Abelian gauge theory.
 
Fortunately, this problem can be avoided by introducing a scalar field that breaks the gauge symmetry spontaneously with its vacuum expectation value $v$ and gives a mass to the gauge field. In this case the contribution to the partition function involves an additional factor $\sim\text{e}^{-(\rho v)^2}$ so that large instantons are exponentially suppressed and the $\rho$-integral becomes finite. Now, performing the $\rho$-integration in (\ref{instcont}) with the exponential suppression factor, dividing by $d^4x$, and ignoring the phase $\text{e}^{\text{i}\theta}$ gives for the instanton density at leading order
 \begin{equation}
  n=Kv^4\text{e}^{-S}\,,
 \end{equation}
 where we have chosen $\rho_0=1/v$ and 
 \begin{equation}
  K\sim C(N)S^{2N}\Gamma\left(\frac{11}{6}N-2\right)
 \end{equation}
The size of the instantons is now effectively cut off at $\rho\sim1/v$ and therefore, as long as $K\text{e}^{-S}\ll1$\footnote{This can always be achieved by choosing the gauge coupling small at the symmetry breaking scale.}, the dilute gas approximation is valid. In particular, the cutoff of the resulting effective theory is given by $v$ as we will treat everything (in particular instantons) smaller than $1/v$ as point-like.
 
 Now we are in a position to sum the contribution of all possible ways to place instantons and anti-instantons in spacetime and find
 \begin{align}
 Z[\theta] & =\sum_{n,\overline{n}=0}^{\infty}\frac{1}{n!\overline{n}!}\prod_{k=1}^{n}\left(\int d^4x_{k}\,Kv^4 e^{-S}e^{\text{i}\theta}\right)\prod_{\overline{k}=1}^{\overline{n}}\left(\int d^4x_{\overline{k}}\,Kv^4 e^{-S}e^{-\text{i}\theta}\right) \nonumber \\
 & =\exp(2K\text{e}^{-S}\int d^4x\,v^4\cos\theta))\,. \label{instpartfunc}
 \end{align}

In the next step we consider YM theory with $N_\text{f}$ Dirac fermions of mass $m$ in the fundamental representation of SU($N$). The corresponding action is\footnote{For the sake of simplicity we have not included the Higgs sector in the action which is, nevertheless, always implicitly assumed to be present.}
\begin{equation}
  S_\text{E}[A_1,\psi,\theta]=\int\frac{1}{2g^2}\text{tr}(F_2\wedge*F_2)-\frac{\text{i}\theta}{8\pi^2}\text{tr}(F_2\wedge F_2)+\sum_{i=1}^{N_\text{f}}\overline{\psi}_i(\hat{\gamma}_\mu\hat{D}_\mu+m)\psi_i*1\,, \label{instfermiaction}
\end{equation}
where $\hat{D}_\mu$ is the Euclidean covariant derivative and $\hat{\gamma}_\mu$ are matrices in spinor space satisfying the Euclidean version of the Clifford algebra, $\{\hat{\gamma}_\mu,\hat{\gamma}_\nu\}=2\delta_{\mu\nu}$. 

Once again we would like to obtain an effective action that is valid below the scale $v$. If $m\gg v$, we can first integrate out the fermions in an instanton background which will give an effective action for the gauge field that has additional terms suppressed by powers of $v/m$. Ignoring these small corrections we are left with a pure gauge theory, i.e. for $v\ll m$ we can simply ignore the fermions at scales below $v$ and the analysis presented at the beginning of this section applies.

For fermions which are light, i.e.~$m\lesssim v$, we can no longer simply integrate them out but have to take them into account properly. For later use let us define the operator-valued matrix $\Psi_{ij}(x)=\overline{\psi}_i(x)P_\text{L}\psi_j(x)$ where $P_\text{L,R}$ denotes the left-handed and right-handed chirality projector, respectively. We would like to integrate out the gauge field and find the effective action for the fermions in a background of a dilute instanton gas. The corresponding calculation has been done by 't Hooft \cite{tHooft:1976snw} (see also \cite{tHooft:1986ooh}). In particular he showed that the partition function corresponding to the action (\ref{instfermiaction}) in an instanton background gives rise to the same fermion propagators as the partition function
\begin{align}
 \int\mathcal{D}\psi\mathcal{D}\overline{\psi}\exp\left(-\int d^4x\sum_{i=1}^{N_\text{f}}\overline{\psi}_i(\hat{\gamma}_\mu\partial_\mu+m)\psi_i\right)
   \frac{d^4zd\rho}{\rho^5}C'(N,N_\text{f})f(g(\rho),N)\text{e}^{\text{i}\theta}\rho^{3N_\text{f}}\text{det}\Psi  \,, \label{instfermiint}
\end{align}
where $z$ and $\rho$ denote position and size of the instanton, $C'$ depends on $N$ and $N_\text{f}$ and, most importantly, the running coupling now includes a contribution due to the fermions:
\begin{equation}
 \frac{8\pi^2}{g^2(\rho)}=\frac{8\pi^2}{g^2(\rho_0)}+\left(\frac{2}{3}N_\text{f}-\frac{11}{3}N\right)\ln\left(\frac{\rho}{\rho_0}\right)+\mathcal{O}\left[g^2(\rho_0)\ln\left(\frac{\rho}{\rho_0}\right)\right]\,. \label{runningcouplingfermions}
\end{equation}
Note that the color-structure of the operator $\text{det}\Psi$ is in general non-trivial but has been suppressed by our simplified notation. In the following the details of this will not be relevant for us but they can be found for example in \cite{tHooft:1976snw,Shifman:1979uw}.

The contribution of an anti-instanton at the same position and of the same size is the complex conjugate of (\ref{instfermiint}). Now we can perform the $\rho$-integration in \eqref{instfermiint} and sum the instanton and anti-instanton contribution as in (\ref{instpartfunc}) to get the following partition function for the fermions
\begin{align}
  Z[\theta]= & \int\mathcal{D}\overline{\psi}\mathcal{D}\psi\exp\left(-\int d^4x\left[\sum_{i=1}^{N_\text{f}}\overline{\psi}_i(\hat{\gamma}_\mu\partial_\mu+m)\psi_i\right.\right. \label{instpartfuncfermi} \\
  & \left.\left.+\kappa v^{4-3N_\text{f}}\text{e}^{-S}(\text{det}\Psi\text{e}^{\text{i}\theta}+(\text{det}\Psi)^\dagger\text{e}^{-\text{i}\theta})\right]\right)\,, \nonumber 
\end{align}
with 
\begin{equation}
 \kappa\sim C'(N,N_\text{f})S^{2N}\Gamma\left(\frac{11}{6}N+\frac{7}{6}N_\text{f}-2\right)\,. \label{kappa}
\end{equation}
From this we see explicitly that integrating out the gauge fields yields effective fermion interactions, also called 't Hooft interactions, which in the case of one flavor reduce to a simple mass term that explicitly reads
\begin{equation}
 \mathcal{L}_\text{mass}=\kappa v\text{e}^{-S}(\overline{\psi}P_\text{L}\psi\text{e}^{\text{i}\theta}+\overline{\psi}P_\text{R}\psi\text{e}^{-\text{i}\theta})\,.
\end{equation}

Next we would like to determine the vacuum expectation value of $\langle\text{tr}(F\wedge F)\rangle$. To do so we need to integrate out the fermions in \eqref{instpartfuncfermi} to obtain an explicit expression for the partition function $Z$ as a functional of $\theta$. The result can be organized in a series expansion in the small quantity $\text{e}^{-S}$ where terms of order $(\text{e}^{-S})^n$ correspond to $n$-instanton contributions. If one is only interested in the leading term of this expansion, one can skip the derivation of the effective action \eqref{instpartfuncfermi} and instead directly integrate out gauge fields and fermions in \eqref{instfermiaction} in one step.

This calculation has also been done by 't Hooft \cite{tHooft:1976snw} and the result differs from that without fermions only by an additional factor $(\rho m)^{N_\text{f}}$ in the instanton contribution (\ref{instcont}) and the proper running coupling as stated in \eqref{runningcouplingfermions}. Furthermore, the constant $C(N)$ in \eqref{instcont} is changed by a factor $\sim\text{e}^{N_\text{f}}$ and hence also depends on $N_\text{f}$ now. Note that the negative contribution $-(2/3)N_\text{f}$ from the running coupling to the exponent of $\rho$ is over-compensated by the factor $(\rho m )^{N_\text{f}}$ so that the $\rho$-integration remains UV finite for all values of $N_\text{f}$.

After repeating the familiar steps of summing all instanton contributions we find
\begin{equation}
 Z[\theta]=\exp\left(2VK'\text{e}^{-S}\left(\frac{m}{v}\right)^{N_\text{f}}v^4\cos\theta\right)\, \label{instpartfuncfermict}
\end{equation}
with
\begin{equation}
 K'\sim C(N,N_\text{f})S^{2N}\Gamma\left(\frac{11}{6}N+\frac{1}{6}N_\text{f}-2\right)\,. \label{kprime}
\end{equation}
Remember that the exponent of this formula is only exact up to order $\text{e}^{-S}$ as we have ignored multi-instanton contributions. Compared to the theory without or with heavy fermions, (\ref{instpartfunc}), we essentially get a suppression factor $(m/v)^{N_\text{f}}$. With this formula it is easy to calculate
\begin{equation}
\langle\text{tr}(F_2\wedge F_2)\rangle=\frac{1}{\text{i}Z[\theta]}\frac{1}{V}\frac{\partial Z[\theta]}{\partial\theta}=16\pi^2\text{i}K'\text{e}^{-S}v^4\left(\frac{m}{v}\right)^{N_\text{f}}\theta
\end{equation}
for small $\theta$ and formally infinite volume $V$ of the 4-dimensional Euclidean space. We observe that the vacuum expectation value vanishes for $m=0$. This is of course the well-known result that massless fermions screen the topological susceptibility of non-Abelian gauge theories. Ultimately, this is due to the fact that massless fermions render $\theta$ unphysical.

\subsection{Forces on domain walls in a dilute instanton gas}

\label{app::b2}

Similarly to the discussion of the 3-form gauge theory in Appendix \ref{app::a} we now want to introduce domain walls via the external source $\theta$ and calculate the forces acting on them. From the effective theory \eqref{instpartfuncfermi} we have seen that instantons induce $2N_\text{f}$-fermion interactions. This implies that fermions can be exchanged between two distinct instantons and hence induce interactions between them. This effect should contribute to the force between domain walls. 

In order to calculate this effect we can determine the vacuum energy $E_0$ in the presence of two domain walls, as defined by \eqref{thetawall}, in perturbation theory and then differentiate it with respect to the distance between the domain walls. For simplicity we consider the case $N_\text{f}=1$ for which the instanton-induced fermion interaction is just a correction to the mass term. We organize our calculation as an expansion in two small parameters. First we assume $\theta$ to be small and keep only terms up to quadratic order in the interaction term in (\ref{instpartfuncfermi}) which gives the following interaction Lagrangian
\begin{equation}
 \mathcal{L}_\text{int}=\tilde{v}\overline{\psi}\left(1-\text{i}\gamma_5\theta-\frac{1}{2}\theta^2\right)\psi\,, \label{efflag}
\end{equation}
where we have introduced $\tilde{v}=\kappa v\text{e}^{-S}$. Our final result, the force on the domain walls, will be given up to quadratic order in $\theta$ as well. Second, there is a factor $\text{e}^{-S}$ in front of the instanton induced interaction term in (\ref{efflag}) which is a measure for the interaction strength. This is a small quantity and therefore we use it as our second expansion parameter. Recall that each instanton comes with this factor and hence we can view terms of order $(\text{e}^{-S})^n$ as an $n$-instanton effect. We will include contributions up to order $(\text{e}^{-S})^2$.

To calculate $E_0$, consider the vacuum to vacuum transition amplitude
\begin{equation}
 \int\mathcal{D}\phi\exp\left(-\int_{-T/2}^{T/2}\mathcal{L}(\phi)\right)\sim\langle0|\text{e}^{-HT}|0\rangle\sim\text{e}^{-E_0T}\,. \label{groundenergy}
\end{equation}
For $T\rightarrow\infty$ and allowing also for a non-zero, static source ($\mathcal{L}(\phi)\rightarrow\mathcal{L}(\phi,J)$), equation \eqref{groundenergy} may be viewed as defining a partition function $Z[T,J]$. Thus, $E_0$ is given by
\begin{equation}
E_0=\lim_{T\rightarrow\infty}-\frac{1}{T}\ln(Z[T,J])+\text{const.}\,, \label{vacenergy}
\end{equation}
i.e.~by the sum of all connected vacuum diagrams in the presence of the source $J$ as claimed in the last paragraph.

Using this result we are in a position to calculate $E_0$ up to second order in $\theta$ and $\text{e}^{-S}$. The two relevant diagrams are shown in Figure \ref{loopdiag}. Making use of the interaction vertex \eqref{efflag} and the $\theta$ profile of \eqref{thetawall} one finds
\begin{align}
E_0=\ldots & +2L^2\left(\tilde{v}\int\frac{d^4p}{(2\pi)^4}\frac{m}{p^2+m^2}-\tilde{v}^2\int\frac{d^3p}{(2\pi)^3}\frac{m^2}{\omega_p^3}\right)((L-(b-a))\theta_1^2+(b-a)\theta_2^2) \nonumber \\
& -L^2\tilde{v}^2\int\frac{d^3p}{(2\pi)^3}\frac{1}{\omega_p^2}\left(\text{e}^{-2\omega_p(b-a)}-1\right)(\theta_2-\theta_1)^2\,, \label{vacenfermi}
\end{align}
where the dots denote $\theta$-independent terms which therefore are irrelevant for the discussion of domain wall effects. Here, as usual, $\omega_p=\sqrt{m^2+p^2}$ and $L^3$ is the formally infinite spatial volume which we assume to be larger than any other scale in the problem. The term linear in $\tilde{v}$ is due to the leading order diagram (cf. Figure \ref{loopdiag1}) which is essentially a propagator in position space evaluated at the origin and integrated over the whole space. All other terms are due to the next-to-leading order diagram (cf. Figure \ref{loopdiag2}) which has a more complicated structure. It consists of two position space propagators evaluated at the difference of two space points $x$ and $y$ over which we have to integrate. Writing the propagators in momentum space gives us two 4-dimensional momentum integrals. We can perform three position integrals which correspond to the directions parallel to the domain walls to obtain a $\delta$-function in momentum space. Consequently, three momentum integrals are trivially performed. The left-over 1-dimensional momentum integral can be performed using the residue theorem. Now we are left with one 4-dimensional position and one 4-dimensional momentum integral as well as the integral over the direction orthogonal to the domain walls. Once again we can perform a 1-dimensional momentum integration using the residue theorem. Although tedious, the leftover position integrals are straightforwardly carried out using \eqref{thetawall}.

Note that, except for the last integral, all momentum integrals are divergent. However, since the effective theory \eqref{instpartfuncfermi} is valid only up to the scale $v$, those integrals are naturally cutoff at this scale. In this case we expect the leading terms, i.e.~those linear in $\tilde{v}$\footnote{Note that there are two terms linear in $\tilde{v}$ of which only one is displayed in \eqref{vacenfermi} since the other term does not depend on $\theta$.}, to parametrically reproduce the vacuum energy that is obtained from the partition function \eqref{instpartfuncfermict} derived by 't Hooft.\footnote{We expect a parametric match only because 't Hooft does not cut off his calculation at $v$ but takes into account all momenta. In contrast, the EFT defined by \eqref{instpartfuncfermi} does not know about physics above $v$.} Since the relevant integral is quadratically divergent this is indeed the case.

Differentiating (\ref{vacenfermi}) with respect to $a$, dividing by $L^2$ and multiplying by $-1$ gives the force density acting on the domain wall at $a$:
 \begin{align}
 f^{(a)} & =2\tilde{v}\left(\int\frac{d^4p}{(2\pi)^4}\frac{m}{p^2+m^2}-\tilde{v}\int\frac{d^3p}{(2\pi)^3}\frac{m^2}{\omega_p^3}\right)(\theta_2^2-\theta_1^2) \nonumber \\ 
 & +2\tilde{v}^2\int\frac{d^3p}{(2\pi)^3}\frac{1}{\omega_p}\text{e}^{-2\omega_p(b-a)}(\theta_2-\theta_1)^2\,. \label{ymforce}
 \end{align}
The second term in the first line simply provides a higher order correction to the leading contribution which we could have alternatively obtained from \eqref{instpartfuncfermict}. However, we also find a completely new contribution to the force at the 2-instanton level proportional to $(\theta_1-\theta_2)^2$. This new term is exponentially suppressed in the distance of the two domain walls.

\bibliography{if}
\bibliographystyle{JHEP}
\end{document}